\documentclass[aps,prb,superscriptaddress,reprint,floatfix,nofootinbib,amsmath,amssymb]{revtex4-2}

\usepackage{graphicx}
\usepackage{dcolumn}
\usepackage{bm}
\usepackage{hyperref}
\usepackage{amsthm}
\usepackage{amsmath}
\usepackage{amssymb}
\usepackage{physics}
\usepackage{enumitem}
\usepackage{textcomp}
\usepackage{amsthm}
\usepackage[capitalize]{cleveref}
\usepackage[caption=false]{subfig}

\newtheorem{theorem}{Theorem}

\theoremstyle{definition}
\newtheorem{lemma}{Lemma}
\newtheorem{prop}{Proposition}
\newtheorem{remark}{Remark}

\newtheorem{definition}{Definition}

\begin{document}

\title{Local Basis Transformation to Mitigate Negative Sign Problems}

\author{Keisuke Murota}
\affiliation{Department of Physics, The University of Tokyo, Tokyo 113-0033, Japan}

\author{Synge Todo}
\affiliation{Department of Physics, The University of Tokyo, Tokyo 113-0033, Japan}
\affiliation{Institute for Physics of Intelligence, The University of Tokyo, Tokyo 113-0033, Japan}
\affiliation{Institute for Solid State Physics, The University of Tokyo, Kashiwa, 277-8581, Japan}

\begin{abstract}

Quantum Monte Carlo (QMC) methods for the frustrated quantum spin systems occasionally suffer from the negative sign problem, which makes simulations exponentially harder for larger systems at lower temperatures and severely limits QMC's application across a wide range of spin systems.
This problem is known to depend on the choice of representation basis.
We propose a systematic approach for mitigating the sign problem independent of the given Hamiltonian or lattice structure.
We first introduce the concept of negativity to characterize the severity of the negative sign problem.
We then demonstrate the existence of a locally defined quantity, the L1 adaptive loss function, which effectively approximates negativity, especially in frustration-free systems.
Using the proposed loss function, we demonstrate that optimizing the representation basis can mitigate the negative sign. This is evidenced by several frustration-free models and other important quantum spin systems.
Furthermore, we compare the effectiveness of unitary transformations against the standard orthogonal transformation and reveal that unitary transformations can effectively mitigate the sign problem in certain cases.
\end{abstract}

\maketitle

\section{\label{sec:intro}Introduction}

Quantum Monte Carlo (QMC) methods are powerful computational tools for studying quantum many-body systems~\cite{Suzuki1994,Avella2013StronglyCS,gubernatis2016quantum}.
The QMC methods map a \( d \)-dimensional quantum system onto a \( (d+1) \)-dimensional classical counterpart via the Suzuki-Trotter decomposition, which allows for the effective simulation of quantum spin systems.
The QMC method is particularly advantageous because it can handle large, strongly correlated systems with high computational efficiency, scaling linearly with the number of lattice sites and the inverse temperature in ideal cases.

Among various QMC algorithms, the loop algorithm~\cite{Evertz1993_qp,Todo2013} and its generalization, 
the worm algorithm~\cite{Prokofev1998, syljuasen2002quantum}, are noteworthy.
The worm algorithm, in particular, is based on the stochastic series expansion (SSE)~\cite{Sandvik1991_yu}
 and is designed to handle the complexities of quantum spin systems by creating and manipulating loops within the lattice.
This method improves the efficiency of sampling configurations and has become a robust, standard approach in QMC simulations.

Despite the advantages of QMC methods, they are significantly hindered by the negative sign problem.
To be more specific, in the case of quantum spin systems, the negative sign problem arises when the Hamiltonian matrix involves positive off-diagonal elements.
This notorious problem makes the direct application of QMC sampling impossible in many intriguing cases, as the Boltzmann weight becomes negative and can not be interpreted as a (unnormalized) probability.
To address this issue, a common strategy is simulating a system with a well-behaved virtual Hamiltonian that closely resembles the original one but has no positive off-diagonal elements and then compensating for the difference through the average sign~\cite{Nakamura1992,Hatano1994,Nakamura1998}.
Approximately, the average sign is given by \(\expval{S} \propto e^{-\beta \Delta F}\), 
where \( \beta \) is the inverse temperature and \( \Delta F \) is the difference in the free energy between the original and virtual systems.
Thus, the sampling efficiency degrades exponentially with the number of lattice sites and the inverse temperature, making simulations increasingly difficult for larger systems at lower temperatures.

So far, several approaches have been explored to alleviate this problem.
One approach focuses on the selection of better virtual Hamiltonians.
In Refs.~\onlinecite{Nakamura1992,Nakamura1998,Hatano1994}, the authors argued that the condition for good virtual Hamiltonians is that \(\Delta F\) is small, meaning the average sign \(\expval{S}\) is close to 1, and the virtual Hamiltonians are constructed based on this strategy.
On the other hand, most modern approaches rely on the representation basis dependencies of the negative sign problem and try to find a negative-sign-free basis for conducting QMC simulations~\cite{Hangleiter2020, Pashayan2015,HatanoSuzuki1992,Levy2021, Nakamura1997, Nakamura1997_pv, Wessel2017,Wessel2018, Okunishi2013,Marvian2019_hh, Klassen2020_en}.
Actually, there are several known examples that completely cure the negative sign problem with this approach, including canonical cluster basis~\cite{Alet2015}, and dimer-like basis rotations~\cite{Okunishi2013, Wessel2017, Miyahara1999, Nakamura1997, Nakamura1997_pv, Wessel2018, Alet2015, Kohshiro2021_jj, Honecker2016_kl}.
Although these methods can eliminate the negative sign problem, they are model-specific and rely on physical intuition to construct the negative-sign-free basis, making them difficult to formulate systematically.
In response to these issues, the current trend is to systematically mitigate the negative sign by optimizing local orthogonal basis transformation with respect to certain loss functions~\cite{Hangleiter2020, Levy2021, Marvian2019_hh, Klassen2020_en}.
These approaches take full advantage of computational power, offering the benefit of reducing the negative sign problem without restricting the target physical systems.

In this paper, we first introduce the concept of ``negativity'' to characterize the severity of the negative sign problem and formulate the optimization problem of the negative sign using this negativity as a loss function.
By analyzing this quantity, we show that the conventional strategy for choosing the virtual Hamiltonian, that is, simply using the absolute value of the weight of the original Hamiltonian, is always the best in this formulation, reducing the problem to simply the optimization of the representation basis.
Ideally, there should be no constraints on the basis transformation, but practically, the transformation matrix should be limited to those represented as a product of local transformations.
Otherwise, the optimization becomes infeasible.
Therefore, in the latter part, we discuss approaches to optimize local transformations using the negativity as the loss function.

In particular, we introduce the \textit{L1 adaptive loss function} as a computationally efficient approximation of the negativity to optimize the local basis instead of directly optimizing the negativity itself, which is computationally intractable.
We also show that the L1 adaptive loss function approximates the negativity well in so-called \textit{frustration-free} quantum spin systems and gives an upper bound.
This fact highlights the importance of the frustration-free property in mitigating the negative sign problem, as seen in the fact that most of the known quantum spin systems with negative-sign-free basis become frustration-free in some parameter regime~\cite{Wessel2018, Wessel2017, Okunishi2013, Miyahara1999, Nakamura1997_pv}.\footnote{It should be noted that frustration-free models are not always free of the negative sign problem. The details will be discussed later.}

Most previous studies on local basis transformations have focused only on orthogonal transformations.
However, unitary transformations can also be utilized in QMC methods.
In this study, we also explore whether unitary transformations have the potential to mitigate the negative sign problem further.

The outline of this paper is as follows: In \cref{sec:qmc}, we review the negative sign problem in the QMC methods and the cause of the negative sign problem.
In \cref{sec:negativity}, we discuss quantifying the negative sign problem by introducing the concept of negativity.
In \cref{sec:local_basis_transformation}, we introduce the L1 adaptive loss function as a good approximation of the negativity and discuss the optimization problem of the local basis transformation using this quantity.
In \cref{sec:num}, we present numerical results demonstrating the effectiveness of our approach on the frustration-free quantum spin models as well as for the various non-frustration-free models.
After that, in \cref{sec:orth_vs_unitary}, we compare the performance of the orthogonal and unitary transformations.
Finally, in \cref{sec:conclusion}, we summarize our findings and potential directions for future research.

\section{Quantum Monte Carlo with Negative Weight}\label{sec:qmc}
First, we review the QMC method and the conventional approach when the Hamiltonian contains positive off-diagonal elements.
In the formalism of SSE, the partition function is calculated by summing weights of all possible configurations of the extended $(d+1)$-dimensional classical system.
Let's assume our system Hamiltonian can be expressed as $H = \sum_{b \in \mathcal{B}} h_b$.
Here, $h_b$ is a local Hamiltonian acting on a single bond $b$, and $\mathcal{B}$ is the set of all bonds in the system.
In the following, we use \(G \equiv -H\) and \(g_b \equiv -h_b\) for \(b \in \mathcal{B}\) to simplify the notation.
Then, the partition function is represented as
\begin{equation}
	\label{eq:partition_function}
	\begin{split}
	Z &= \text{Tr}\,[\exp(\beta G)] 
		\approx \sum_{\psi_1} \bra{\psi_1} \prod_{k=1}^{M} ( 1 +  \frac{\beta}{M}G ) \ket{\psi_1} \\
	   &= \sum_{\{\psi_k\}} \bra{\psi_1} ( 1 + \frac{\beta}{M}G ) \ket{\psi_2} \cdots \bra{\psi_M} ( 1 + \frac{\beta}{M}G ) \ket{\psi_1} \\
	   &= \sum_{\{\psi_k\}} \sum_{\{b_k\}} \bra{\psi_1} ( 1 + \frac{\beta}{M} g_{b_1} ) \ket{\psi_2} \cdots \\
	   & \qquad \times \bra{\psi_M} ( 1 + \frac{\beta}{M} g_{b_M} ) \ket{\psi_1}.
	\end{split}
\end{equation}
Here, $\ket{\psi}$ is the basis of the Hilbert space of the system and 
the collections of states and bonds, $\psi_1, \dots, \psi_M$ and $b_1, \dots, b_M$, define a single configuration of the system, $c_M$.
Note that in the actual QMC simulation, we can rigorously take the infinite $M$ limit, and the simulation becomes unbiased,
thus, we will express the configuration without the label $M$ to indicate the infinite $M$ limit in the following.
Finally, by expressing the weight of a configuration $c$ as $W(c)$,
we can write the partition function as the sum of weights of all configurations:
\begin{equation}
    Z = \sum_{c \in \mathcal{C}} W(c).
\end{equation}
Here, $\mathcal{C}$ is the set of all configurations of the system.

Usually, with above definition of $W(c)$, the expectation of an observable $\mathcal{O}$ can be expressed as 
\begin{equation}
    \label{eq:obs}
    \expval{\mathcal{O}} = \frac{1}{Z} \sum_{c \in \mathcal{C}} W(c) \mathcal{O}(c).
\end{equation}
In QMC, \cref{eq:obs} is interpreted as sample average over distribution $W(c)/\sum_{c \in \mathcal{C}} W(c)$ as long as all weights are non-negative.
On the flip side, the situation becomes exponentially more complex in the case of $W(c)$ taking negative values.
In this case, $W(c)/Z$ cannot be regarded as a probability distribution, 
and the so-called reweighting approach is used as a workaround.
More concretely, instead of simulating the Hamiltonian $G$, we simulate a virtual Hamiltonian $G_\text{v}$, where off-diagonal elements are all positive and then reweight the results afterward as follows:
\begin{equation}
    \label{eq:vham}
    \begin{split}
        \expval{\mathcal{O}} =& \frac{
            \displaystyle \sum_{c\in \mathcal{C}} W_\text{v}(c) \frac{W(c)}{W_\text{v}(c)} O(c)
        }
        {
            \displaystyle \sum_{c\in \mathcal{C}} W_\text{v}(c) \frac{W(c)}{W_\text{v}(c)} 
        }
        = \frac{\expval{S O}_\text{v}}{\expval{S}_\text{v}}.
    \end{split}
\end{equation}
Here, $\expval{A}_\text{v}$ denotes the expectation value of $A$ evaluated for the virtual Hamiltonian $G_\text{v}$.
We also defined the reweighting factor $S(c) = W(c)/ W_\text{v}(c)$.
The expectation value of the reweighting factor is often called the \textit{average sign}.

\section{Quantification of negative sign problem}
\label{sec:negativity}
Simple analytical calculation of the statistical error of the estimator \cref{eq:vham} tells that temperature dependency of the statistical error is fully characterized by the relative error of the average sign, $\eta = \sqrt{\expval{|S|^2}_\text{v} / |\expval{S}_\text{v}|^2}$.
In the following, we call this quantity the \textit{negativity}, and by optimizing \(\eta\) either via basis transformation or choice of a virtual Hamiltonian, we try to mitigate the effect of the negative sign problem.
Interestingly, however, as for the choice of the virtual Hamiltonian, we can confirm that the standard choice \(G_\text{v}=G^+\), the so-called stoquastic of \(G\), where each non-diagonal element in \(G\) is replaced by its absolute value, always achieves the optimal negativity as shown in the following proposition:
\begin{prop}[Optimal virtual Hamiltonian]
	\label{prop:opt_v_ham}
	We first express the negativity using the weight $W(c)$ and $W_\text{v}(c)$:
	\begin{equation}
		\label{eq:eta_weight}
	    \begin{split}
		    \eta^2 &= \frac{\expval{|S|^2}_\text{v}}{|\expval{S}_\text{v}|^2} =
			\frac{\displaystyle \sum_{c \in \mathcal{C}} |W(c)|^2 / W_\text{v}(c)}{\displaystyle \sum_{c \in \mathcal{C}} W_\text{v}(c)} \cdot
			\frac{\displaystyle \Big|\sum_{c \in \mathcal{C}} W_\text{v}(c)\Big|^2}{\displaystyle \Big|\sum_{c \in \mathcal{C}} W(c)\Big|^2} \\
		    &= \frac{\displaystyle \Big[\sum_{c \in \mathcal{C}} |W(c)|^2 / W_\text{v}(c)\Big] \cdot \displaystyle \Big[\sum_{c \in \mathcal{C}} W_\text{v}(c)\Big]}{\displaystyle \Big[\sum_{c \in \mathcal{C}} W(c)\Big]^2} \\
		    & \geq 
	        \frac{\displaystyle \Big[\sum_{c \in \mathcal{C}} \sqrt{|W(c)|^2 / W_\text{v}(c)} \cdot \sqrt{W_\text{v}(c)}\Big]^2}{\displaystyle \Big[\sum_{c \in \mathcal{C}} W(c)\Big]^2} \\
		    & = \frac{\displaystyle \Big[\sum_{c \in \mathcal{C}} |W(c)|\Big]^2}{\displaystyle \Big[\sum_{c \in \mathcal{C}} W(c)\Big]^2} = \left[\frac{\Tr\,[\exp(\beta G^+)]}{\Tr\,[\exp(\beta G)]}\right]^2.
	    \end{split}
	\end{equation}
  Here, we used the Cauchy–Schwarz inequality between $\sqrt{W_\text{v}(c)}$ and $\sqrt{|W(c)|^2/W_\text{v}(c)}$ to obtain inequality from the second line to the third line.
  We also use \(W_\text{v}(c) > 0\) \(\forall c\) and \(\sum_{c \in \mathcal{C}} W(c) = Z > 0\).
  The equality on the third line is achieved when $W_\text{v}(c) = |W(c)|$, which is equivalent to $G_\text{v} = G^+$.
\end{prop}

This proposition claims that we do not have to care about the virtual Hamiltonian in the first place.
By choosing $G_\text{v} = G^+$, the negativity reduces to $\eta = |\expval{S}_\text{v}|^{-1}$, i.e., the severity of the negative sign problem is characterized by the average sign.
Similar discussion is also done in Refs.~\onlinecite{Hangleiter2020, Pashayan2015}.

\section{Local Basis Transformation}
\label{sec:local_basis_transformation}

\subsection{Negativity in low temperature limit}

When both $G$ and $G^+$ have unique ground state, the average sign has the following asymptotic form
\begin{equation}
    \label{eq:asympt}
    \begin{split}
    \eta &= |\expval{S}_\text{v}|^{-1} = \frac{
			  \displaystyle \sum_{c \in \mathcal{C}} W_\text{v}(c) 
    }
    {
			  \displaystyle \Big| \sum_{c \in \mathcal{C}} W_\text{v}(c) \cdot \frac{W(c)}{W_\text{v}(c)} \Big|
    } \\
    &=
    \frac{
			  \displaystyle \sum_{c \in \mathcal{C}} W_\text{v}(c) 
    }
    {
			  \displaystyle \sum_{c \in \mathcal{C}} W(c)
    } 
    \approx
    e^{\beta(\lambda^+(G) - \lambda(G))}
    \end{split}
\end{equation}
for large $\beta$.
Here, $\lambda(G)$ and $\lambda^+(G)$ denote the largest eigenvalue of \(G\) and $G^+$, respectively.
Since the average sign approaches zero exponentially for larger $\beta$, it is sometimes more convenient to use $\eta^{\prime} \equiv \lambda^+(G) - \lambda(G)$ as a loss function.
We call this quantity the \textit{gap negativity}.
Since the negative sign problem becomes critical at lower temperatures, the gap negativity can be regarded as a temperature-independent quantity that characterizes its severity.
Note that $\lambda(G)$ is independent of the choice of representation basis.
Therefore, our aim now is to simply minimize the largest eigenvalue of $G^+$.

\subsection{Frustration-Free Systems}\label{sec:frustration_free_system}

\begin{figure}[tbp]
	\centering
	\subfloat[]{
		\includegraphics[width=0.4\textwidth]{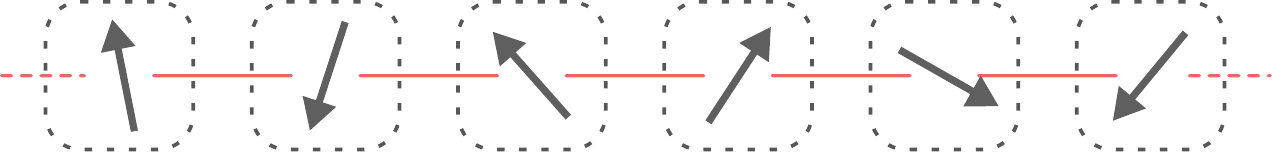}\label{fig:unit_cells_a}
	} \\
	\subfloat[]{
		\includegraphics[width=0.4\textwidth]{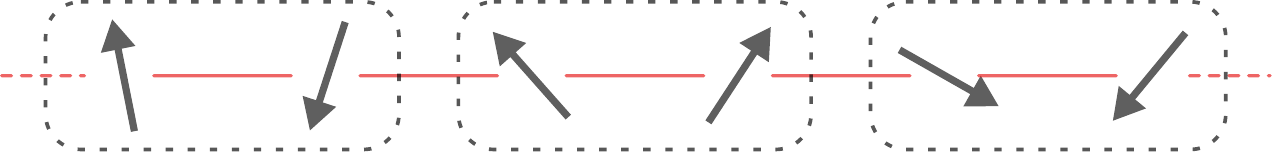}\label{fig:unit_cells_b}
	}
	\caption{Illustration of the unitary transformation and the choice of cells for quantum spin chains.
		Here, we consider two types of cells to which the same local unitary transformation is applied.}\label{fig:unit_cells}
\end{figure}

Although the formulation based on the negativity or the gap negativity is straightforward, the actual computational cost to evaluate \cref{eq:asympt} is exponentially high without further approximation or assumption.
First, as a fundamental and natural requirement, we limit the unitary matrices for the basis transformation to those that can be expressed as a product of local unitary transformations.
Mathematically, this requirement is expressed as
\begin{equation}
    U = \otimes_i u_i,
\end{equation}
where $u_i$ is the local unitary transformation acting on the cell labeled by $i$.
Theoretically, the cell can be chosen arbitrarily, and we can apply different unitary on each cell.
However, in practice, we often select the cells as small repeating units, and the same local unitary transformation is shared across all cells.
An example of the choice of cells is shown in \cref{fig:unit_cells}.

To derive an approximation of the loss function, we additionally assume that the Hamiltonian under consideration possesses the frustration-free property~\cite{Selke1988, Peschel1981, Yu_Kitaev1997_rr, Rokhsar1988,Affleck1988_ps}.
While this assumption might appear overly restrictive at first glance, it is reasonable when considering the improvement of the negative sign problem through local basis transformations.
Indeed, upon re-examining previous studies on analytical local transformation methods, it becomes apparent that the frustration-free property plays a crucial role in all spin models in which local basis transformation can eliminate the negative sign problem.
Thus, for now, we will derive an approximation for the L1 adaptive loss under two assumptions: (1)~local unitary transformations and (2)~the frustration-free property.

The definition of frustration-free property is fairly simple:
\begin{definition}[Frustration-free Hamiltonian]\label{def:ff_ham}
	Consider a Hamiltonian written as $G = \sum_{b\in\mathcal{B}} g_b$, with $e^0_b$ being the ground state energy (or the largest eigenvalue) of $g_b$.
	If the ground state energy of the system is $\sum_{b\in\mathcal{B}} e^0_b$, then the Hamiltonian with this local Hamiltonian decomposition is called \textit{frustration-free}.
\end{definition}
\begin{lemma}[Property of frustration-free Hamiltonian]
	If $G = \sum_{b\in\mathcal{B}} g_b$ is a frustration-free Hamiltonian, then its ground state (or the eigenstate associated with the largest eigenvalue) $\ket{\psi_G}$ is also the ground state of $g_b$, i.e., $g_b \ket{\psi_G} = e^0_b \ket{\psi_G}$.
	Conversely, if there is a state $\ket{\psi}$ such that $g_b \ket{\psi} = e^0_b \ket{\psi}$ for all $b$, $G$ is a frustration-free Hamiltonian and $\ket{\psi}$ is the ground state of $G$. The proof of this lemma is given in Ref.~\onlinecite{Tasaki_undated}.
\end{lemma}
It is important to note that frustration-free property is always defined with respect to the Hamiltonian $G$ and its local Hamiltonian decomposition $G = \sum_{b\in\mathcal{B}} g_b$.

\subsection{L1 Adaptive Loss}

Although this may seem somewhat ad hoc, we define the L1 adaptive loss as follows:
\begin{definition}[L1 adaptive loss]\label{def:local_loss}
	Let $G = \sum_{b \in \mathcal{B}} g_{b}$ be a Hamiltonian and
	$u$ be a local unitary matrix acting uniformly on all sites.
	Then, the \textit{L1 adaptive loss} is defined as
	\begin{equation}
		\label{eq:local_loss}
        \begin{split}
		&\mathcal{L}_{\text{L1}}(G, u) =
		\frac{1}{|\mathcal{B}|} \sum_{b \in \mathcal{B}}
		\lambda^+((u \otimes u) g_b (u^\dagger \otimes u^\dagger))
		- \lambda(g_b) \\
        &= 
        \text{max}_{\ket{\psi}}
        \frac{1}{|\mathcal{B}|} \sum_{b \in \mathcal{B}} \sum_{i, j} \psi_i \psi_j^* 
        \vert {((u\otimes u) g_b (u^\dagger \otimes u^\dagger))}_{ij} \vert  \\
        & \qquad + \text{const},
        \end{split}
	\end{equation}
	where $|\mathcal{B}|$ is the number of bonds in the set of bonds $\mathcal{B}$ and $\ket{\psi}$ is any normalized state of the system, i.e., $\sum_i |\psi_i|^2 = 1$.
	From the Perron-Frobenius theorem~\cite{Perron1907-kq}, as the eigenstate corresponding to the largest eigenvalue of $A^+$ can always be taken as a non-negative valued vector, we can assume that $\ket{\psi}_i$ is also non-negative without loss of generality.
	Embodying this, we can interpret \cref{eq:local_loss} as a Min-Max optimization problem to minimize the L1 norm of the local Hamiltonian $U g U^\dagger$ with each element
	weighted by $\psi_i \psi_j^*$. 
	Since the vector $\ket{\psi}$ changes over time, we call this loss the \textit{L1 adaptive loss}.
	Remember that $\lambda(A)$ is the largest eigenvalue of $A$ and $\lambda^+(A)$ is defined as $\lambda(A^+)$, i.e., the largest eigenvalue of $A^+$.
	Another important property of this loss function is that when the Hamiltonian has translational symmetry, this quantity can be defined independently of the system size.
	This property allows us to optimize local unitary transformation efficiently.
\end{definition}

\begin{remark}\label{rem:stoquastic_local_ham}
	It is important to note that the stoquastic of $G$ is not always expressed in the sum of stoquastic local Hamiltonians.
	This is because stoquastic operation involves taking absolute value element-wisely,	which does not adhere to the distributive law.
	As a result, transforming the entire Hamiltonian into its stoquastic form is not equivalent to simply summing the stoquastic forms of its local components:
	\begin{equation}
		\label{eq:stoquastic_local_ham_neq}
            G^+ \neq \sum_{b \in \mathcal{B}} g_b^+.
	\end{equation}
	Since the inconsistency between the left and right-hand sides in \cref{eq:stoquastic_local_ham} is raised only at the overlapping sites of the local Hamiltonians, the difference between the two	depends on the lattice connectivity, or more precisely, the ratio between area and volume of cells.
	This fact suggests that by taking large cells, the effect of the overlapping sites will be suppressed.
	It is important to note that in cases where the lattice connectivity is restricted to neighbors, we can still express $G^+$ as a sum of stoquastic local Hamiltonians.
	This is represented in the following equation:
	\begin{equation}
		\label{eq:stoquastic_local_ham}
		G^+ = \sum_{n(b) \in \mathcal{B}} {\left(\tilde{g}_{n(b)}\right)}^+.
	\end{equation}
	Here, $n(b)$ denotes the set of all nearest neighbors involved in the bond $b$, and $\tilde{g}_{n(b)}$ is a local Hamiltonian that includes all the off-diagonal elements acting on $b$.
	In other words, it servers as a local Hamiltonian containing all the action supported on $b$.
	For the sake of simplicity, we will replace $g_b$ with this local term in the following discussion.
	A detailed discussion of this remark can be found in the supplementary material of Ref.~\onlinecite{Hangleiter2020}.
\end{remark}

Although the L1 adaptive loss might appear to be insufficient in capturing the essence of the system, 
the following theorem guarantees that it provides a good approximation in the case of frustration-free quantum spin models.

\begin{theorem}\label{thm:local_loss_upper}
	The L1 adaptive loss gives the upper bound of gap negativity:
	\begin{equation}
		\label{eq:local_loss_upper}
		\mathcal{L}_{\text{L1}}(G, u) \geq \lambda^+(G) - \lambda(G).
	\end{equation}
\end{theorem}
\begin{proof}
	Since $G^+$ is a sum of local Hamiltonians $g_b^+$,	we have
	\begin{equation}
        \label{eq:local_loss_upper_proof}
		\lambda^+(G) \leq \sum_{b \in \mathcal{B}} \lambda^+(g_b).
	\end{equation}
	Then, since the Hamiltonian is frustration-free, the following inequality is derived:
	\begin{equation}
		\sum_b \lambda(g_b) - \lambda^+(g_b) \geq \lambda(G) - \lambda^+(G) \geq 0.
	\end{equation}
	 By using \cref{def:local_loss}, we can prove \cref{thm:local_loss_upper}.
\end{proof}

In deriving the L1 adaptive loss, we initially assumed the frustration-free property.
However, even in cases where the Hamiltonian is not strictly frustration-free, this quantity is expected to serve as a good approximation of the average sign, particularly when the system's ground state closely resembles that of the local Hamiltonians.
In other words, the L1 adaptive loss remains a good approximation when the original Hamiltonian has a certain frustration-freeness~\cite{Takahashi2023-te}.
Furthermore, by recognizing that all previously studied loss functions can be expressed as $\sum_{ij} \vert (u\otimes u) g_b (u^\dagger \otimes u^\dagger) \vert_{ij}$, which is essentially the L1 loss of the local Hamiltonians, we can identify a key distinction: 
Previous loss functions assumed that the ground state is a uniform wave function in the transformed basis, whereas the L1 adaptive loss recalculates the ground state at each iteration, hence, providing a more accurate approximation.

\section{Numerical Result}\label{sec:num}
Based on the above discussion,
we will demonstrate through numerical optimization that the negative sign problem can indeed be mitigated.
Before actually examining the research results, let us briefly explain the environmental settings and
approaches used in the numerical calculations.
It should be noted that our optimization method is independent of the specific choice of Markov chain Monte Carlo (MCMC)~\cite{Robert2010-iw,Landau2014-za, Newman1999-ww}
sampling algorithm employed.
In our case, we used the modified worm algorithm
to simulate local degrees of freedom beyond binary values
by accommodating general integer degrees of freedom. 
The transition kernel of the worm algorithm is designed to 
minimize the rejection rate during the worm updating process as 
explained in Ref.~\onlinecite{Suwa2011_wx}.
Additionally, our generalized worm algorithm incorporates warp updates to introduce single-site updates~\cite{Suwa2012}.
Regarding the optimization method using local basis transformations, we employed the Riemannian gradient descent algorithm~\cite{Abrudan2008-lo,Abrudan2009-tp,Lezcano-Casado_undated-xk,Lezcano-Casado_undated-xk2} and used Adam as the optimizer~\cite{Kingma2014-rn}.

This section is structured as follows:
First, we demonstrate a proof of concept by applying our optimization method to artificially constructed one-dimensional (1D) and two-dimensional (2D) frustration-free models.
Following this, we present the results of our method for models studied in previous analytical research, which use local basis transformations based on physical intuition to eliminate the negative sign problem, including the $J_0$-$J_1$-$J_2$-$J_3$ model, bilinear-biquadratic chain, frustrated ladder, and Shastry-Sutherland model.
Finally, we show that it is also possible to mitigate the negative sign problem to some extent in the kagome Heisenberg model, which has strong frustrations.

\subsection{Randomly Generated Frustration-Free Model}

As the derivation of the L1 adaptive loss function, \cref{eq:local_loss}, highly depends on the frustration-free property of the Hamiltonian, we demonstrate the applicability of our optimization scheme to randomly generated Hamiltonians belonging to this class.

\begin{figure}[tbp]
	\centering
	\includegraphics[width=0.4\textwidth]{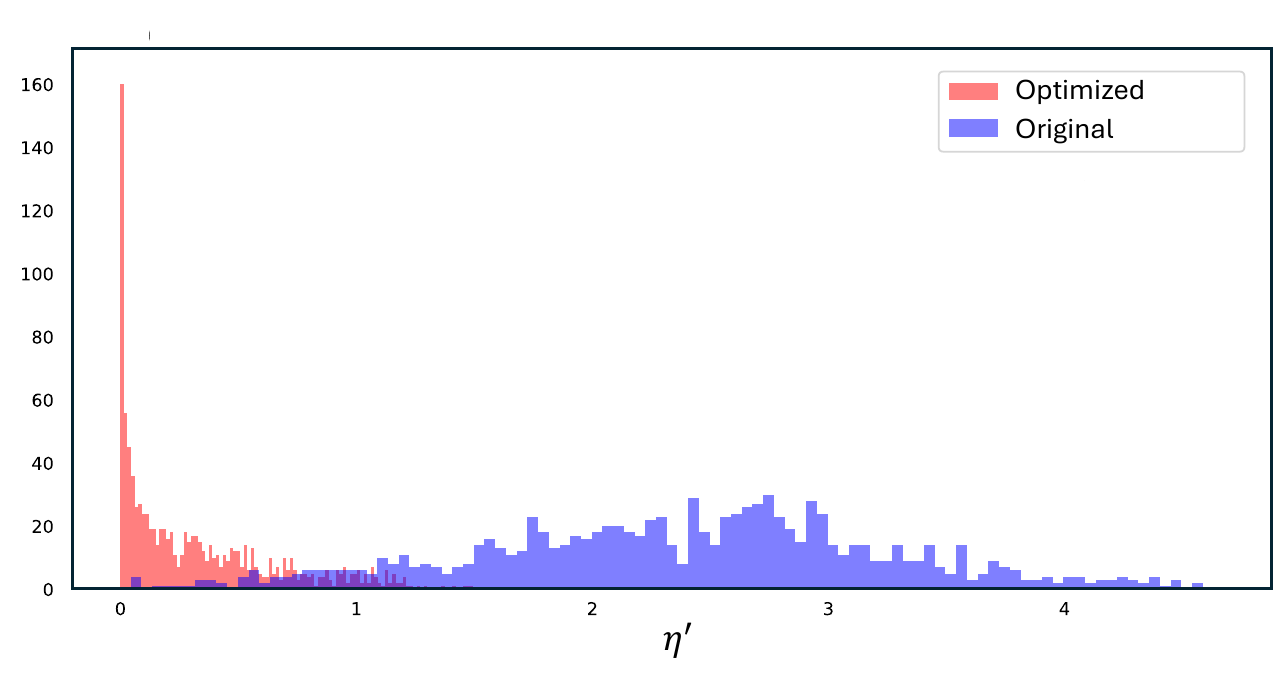}
	\caption{
		Histogram of gap negativity $\eta^\prime$ before and after the optimization
		for 1D random frustration-free models with $L = 6$.
		The blue histogram represents the gap negativity of the original Hamiltonian, and the red histogram represents the gap negativity of the optimized Hamiltonian.
		Optimization is done by minimizing the L1 adaptive loss function for cell type~(a) in \cref{fig:unit_cells}.
		Evaluation of the gap negativity is done by exact diagonalization of the system Hamiltonian and its stoquastic.
}\label{fig:1d_random_ff_6}
\end{figure}

First, we consider frustration-free models defined on a chain lattice.
For simplicity, only the model Hamiltonians with translational invariance and periodic boundary conditions (PBC) will be our primal test cases.
To generate 1D frustration-free models, we exploit the concept of the parent Hamiltonian~\cite{PerezGarcia2006Matrix, Fernandez-Gonzalez2012-eg} 
that can be defined as a matrix product state (MPS)~\cite{Klumper1993-sa}.
In the present simulation, we construct frustration-free models as a parent Hamiltonian of random matrix product state with bond dimension $\chi=2$ defined on chain lattice with three-dimensional local Hilbert space.
By construction, the local Hamiltonian is a projection matrix.
Thus, the loss function can always be in the range $[0, 1]$.
As a benchmark, first, we perform our algorithm on 1,000 random 1D frustration-free models and plot the histogram of the gap negativity before and after optimization in \cref{fig:1d_random_ff_6}.
One can see that the red histogram (optimized) is more concentrated and distributed in lower negativity regions, while the blue histogram (original) is more spread out.
This outcome effectively demonstrates the performance of our algorithm in systematically reducing the negativity in 1D randomly generated frustration-free models.

\begin{figure}[tbp]
	\centering
	\includegraphics[width=0.5\textwidth]{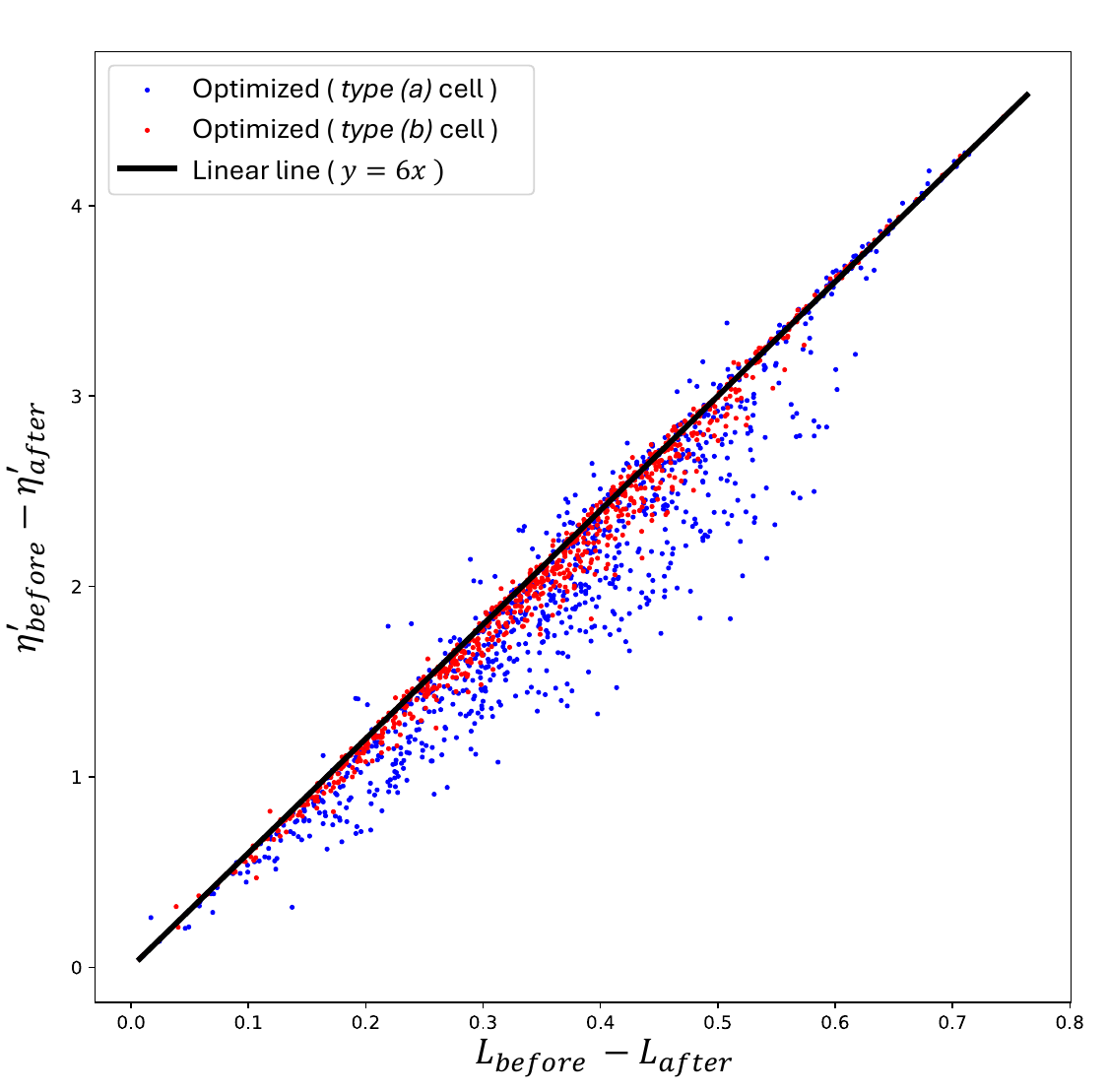}
	\caption{
		Scatter plot of improvements in gap negativity (vertical axis) and L1 adaptive loss function (horizontal axis) for 1D random frustration-free models with $L = 6$.
		The improvement of gap negativity and loss function is defined as
		$\eta^\prime_{\text{before}} - \eta^\prime_{\text{after}}$ and
		$\mathcal{L}_{\text{before}} - \mathcal{L}_{\text{after}}$, respectively.
		Blue dots show the results using cell type~(a) in \cref{fig:unit_cells} and red dots with cell type~(b).
	}\label{fig:scatter_neg_ff_6}
\end{figure}

Next, \cref{fig:scatter_neg_ff_6} shows the relation between the negativity and the L1 adaptive loss function.
The figure illustrates the change in the negativity and the L1 adaptive loss function before and after the optimization.
The data points in the upper right corner indicate a significant reduction in the L1 adaptive loss function compared to its pre-optimization value and a substantial alleviation of the negative sign in the system Hamiltonian.
A notable characteristic of this plot is the linear correlation observed across
the data, suggesting that negativity acts as a decreasing function of the L1 adaptive loss function.
This linear relationship is indeed a consequence of \cref{thm:local_loss_upper},
in this specific instance with $L=6$.
In the plot, there is a black line representing $y = 6x$.
The coefficient 6 comes from the fact $|\mathcal{B}|=6$.
One can observe most of the points are indeed below this line,
while there are some points exceeding this line.
This happens due to the fact that the local Hamiltonian we optimized here is
not $\tilde{g}^+$ [\cref{eq:stoquastic_local_ham_neq}].
However, this extra negativity will be less influential in cell type~(b) in \cref{fig:unit_cells}, where the unitary transformation is applied to two sites.
One can indeed confirm this by noting that red points are more concentrated below the black line.
This is because the effect of the sign problem existing at the edges of the cells
contributes less to the entire negativity in cell type~(b).
See the discussion in \cref{rem:stoquastic_local_ham}.

\begin{figure}[tbp]
	\centering
	\includegraphics[width=0.45\textwidth]{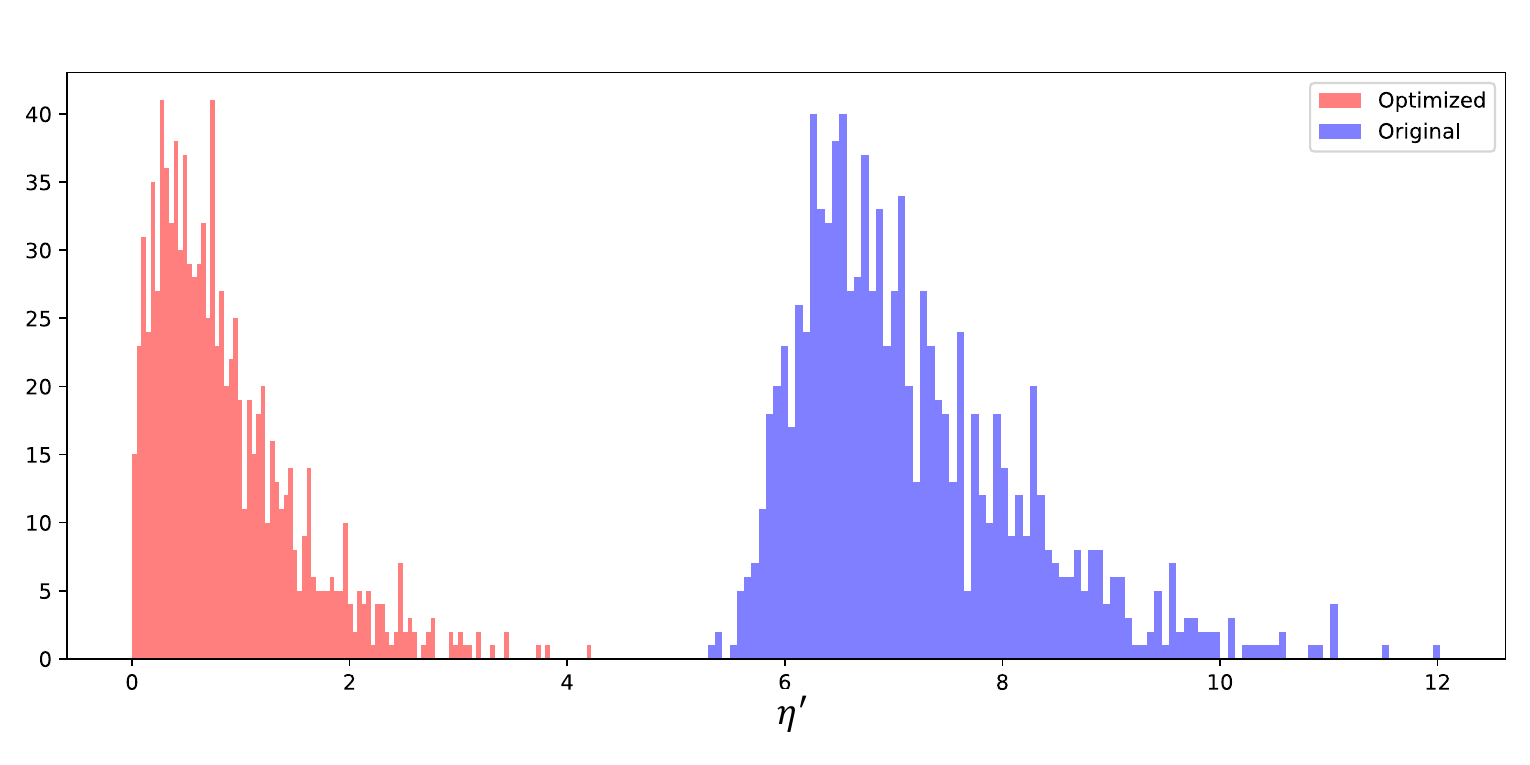}
	\caption{
    Histogram of gap negativity $\eta^\prime$ for original Hamiltonian (blue) and for optimized Hamiltonian (red)
	for 2D random frustration-free models.
	}\label{fig:hist_ff2D}
\end{figure}

We also demonstrate the effectiveness of our method for 2D frustration-free models.
We generate 1,000 Projected Entangled Pair States (PEPS)~\cite{Orus2014-wb} and construct their parent Hamiltonians based on the idea explained in Refs.~\onlinecite{PerezGarcia2007PEPS, Schuch2010PEPS}.
In constructing the parent Hamiltonians, it is necessary to set the bond dimension to at least 8 (it is three in the 1D case).
Consequently, our 2D frustration-free models defined on a square lattice have a local Hilbert space with dimension 8.
A histogram illustrating the extent to which our optimization method has mitigated the negative sign compared to the original 2D frustration-free Hamiltonians is presented in \cref{fig:hist_ff2D}.
As observed in the figure, the gap negativity $\eta^\prime$ after the optimization is significantly reduced compared to $\eta^\prime$ for the original Hamiltonians.
Note $\eta^\prime=0$ means that the negative sign is removed completely.
This reduction of the gap negativity indicates improvement in the negative sign by our method, demonstrating its effectiveness in addressing the negative sign problem of 2D quantum spin systems as well.
Considering that previous studies on the negative sign problem have scarcely reported results on mitigating in 2D models, the present outcome has significant importance.
Indeed, the only previously known result about reducing the sign problem for the 2D case is Ref.~\onlinecite{Wessel2018}, and thus, the present result is a major advance in that it provides a robust methodology for addressing this complex challenge in quantum simulation.

\subsection{Benchmark on Models with Known Negative-Sign-Free Transformations}
There are several frustration-free models for which local basis transformations that eliminate the negative sign are known analytically.
As described in \cref{sec:frustration_free_system}, our proposed loss function is inspired by the frustration-free property.
This background on the design of the loss function suggests that optimizing the L1 adaptive loss function should be able to reconstruct the analytically derived local basis transformations or other optimal transformations similar to it.
In this section, we rigorously examine this postulation by applying our methodology to various quantum spin models and compare with established theoretical predictions.

\begin{figure}[tbp]
  \centering
    \includegraphics[clip,width=0.8\columnwidth]{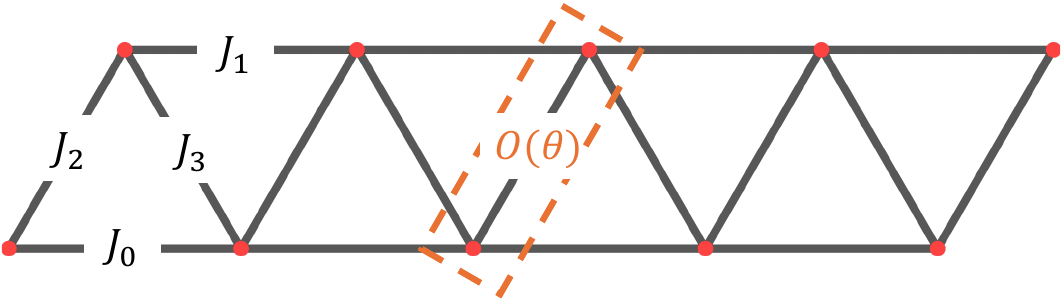}%
    \caption{
      Lattice structure and cell for basis transformation for the $J_0$-$J_1$-$J_2$-$J_3$ model.
      We only consider the case when $J_0 = J_1 = 1$.
    }\label{fig:unit_cell_j0j1j2j3}
\end{figure}

\subsubsection{\texorpdfstring{$J_0$-$J_1$-$J_2$-$J_3$}{J0-J1-J2-J3} Model}
\begin{figure}[tbp]
	\centering
	\includegraphics[clip,width=0.9\columnwidth]{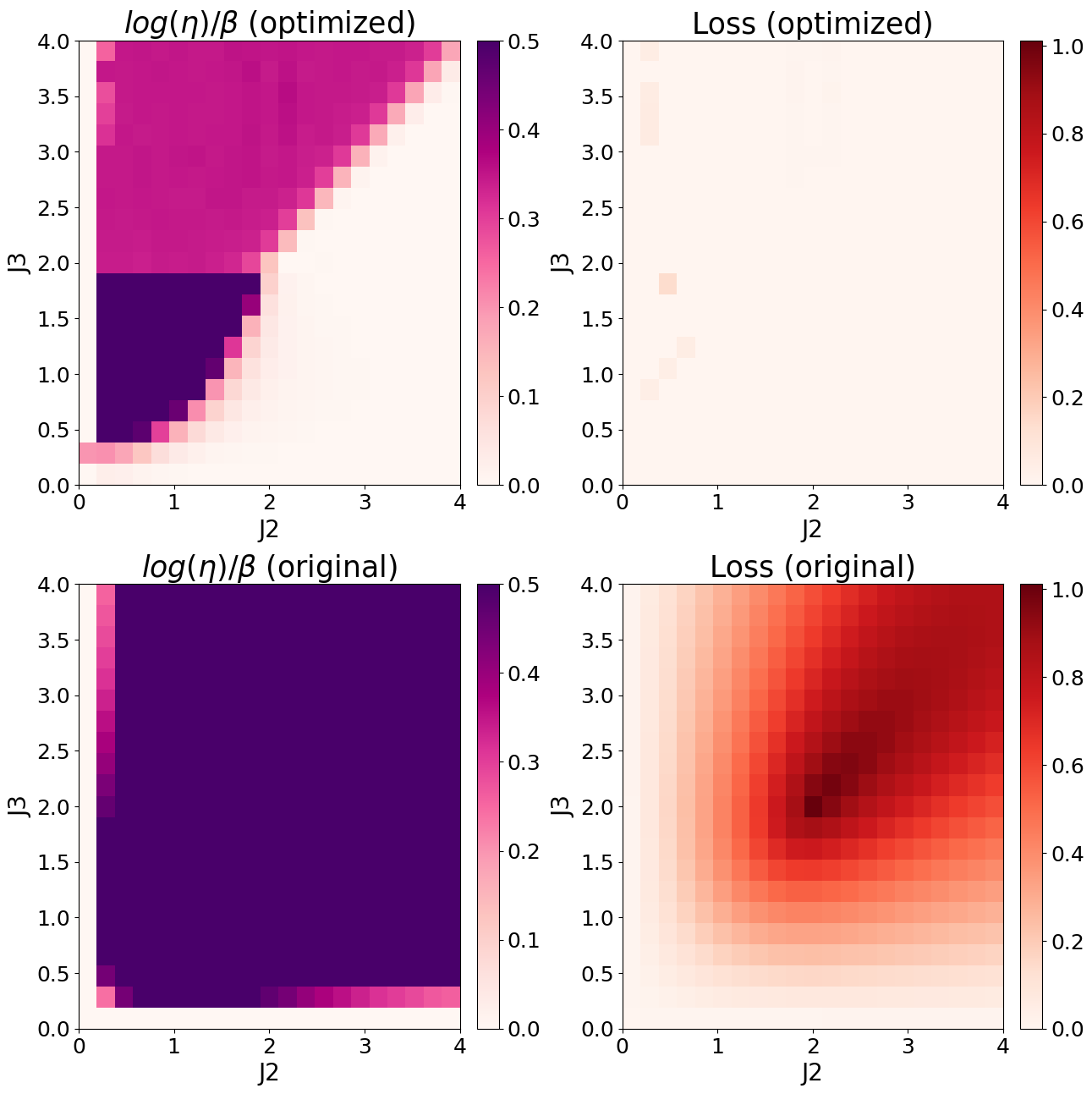}
	\caption{
        Gap negativity $ \eta^\prime \approx \log(\eta) / \beta$ (left) and L1 adaptive loss function (right) before (bottom) and after (top) the optimization for the $J_0$-$J_1$-$J_2$-$J_3$ model at $\beta = 4$.  The next nearest neighbor interactions $J_0$ and $J_1$ are fixed to unity, and the heat map is plotted for various $J_2$ and $J_3$.
        Simulation is done for $L = 10$.
	Color gradients indicate the magnitude of $\log(\eta) / \beta$ or the L1 adaptive loss function: 
	The lighter the color, the better.
	}\label{fig:heat_map_mg1d}
\end{figure}
\begin{figure}[tbp]
	\centering
	\includegraphics[clip,width=0.9\columnwidth]{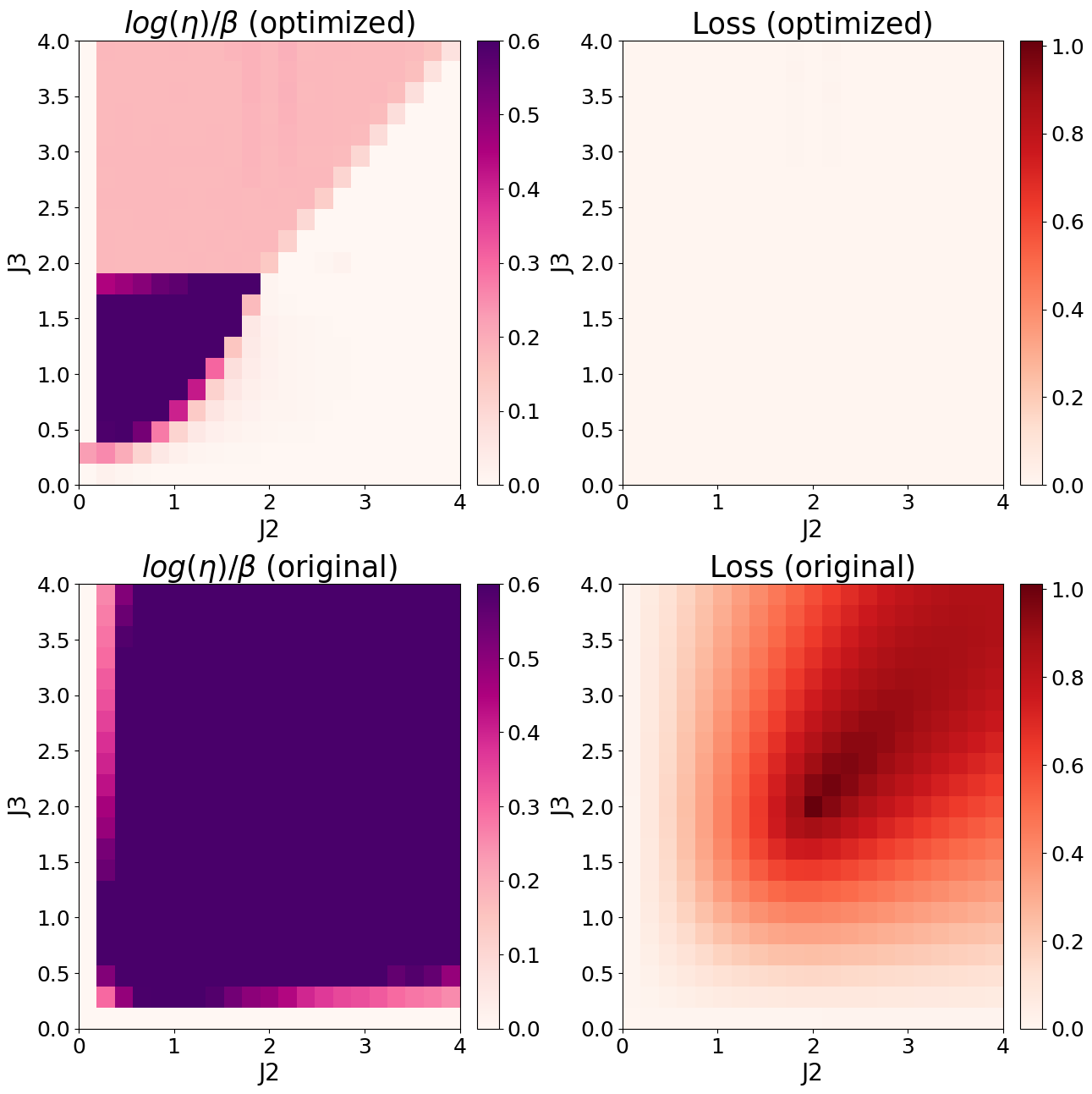}
	\caption{
        Gap negativity $\eta^\prime \approx \log(\eta) / \beta$ (left) and L1 adaptive loss function (right) before (bottom) and after (top) the optimization for the $J_0$-$J_1$-$J_2$-$J_3$ model at $\beta = 8$.
	}\label{fig:heat_map_mg1d_4}
\end{figure}

First, we apply our method to the $J_0$-$J_1$-$J_2$-$J_3$ model, which is a spin-1/2 Heisenberg chain with nearest-neighbor antiferromagnetic interactions $J_2$ and $J_3$ and next-nearest-neighbor antiferromagnetic interactions $J_0$ and $J_1$.
We choose the cells for local orthogonal basis transformation as shown in \cref{fig:unit_cell_j0j1j2j3}.
In the simulation, $J_0$ and $J_1$ are fixed to unity.
The result is shown in \cref{fig:heat_map_mg1d}.
In Ref.~\onlinecite{Nakamura1997_pv}, the authors analytically found that the negative sign can be completely removed when $J_3 > J_1 + J_0 = 2$ with dimer-like basis rotation.
In \cref{fig:heat_map_mg1d}, the negative sign is mitigated for $J_3 < J_2$, while the region $J_3 > 2 \ \cap \ J_3 > J_2$ still appears to be affected by the negative sign.
The reason for the apparent persistence of the negative sign in this region is that, although $\log(\eta) / \beta$ reaches the minimum value after optimization, the degeneracy of the ground state increases compared to the original Hamiltonian $G$.
The reason for the apparent persistence of the negative sign in this region is that, although $\log(\eta) / \beta$ reaches the minimum value after optimization, the degeneracy of the ground state increases compared to the original Hamiltonian $G$.
However, since this degeneracy only affects the negative sign by a constant factor, the average sign does not diverge at low temperatures.
This implies that we are not fundamentally facing the negative sign problem in this region.
This can also be seen from \cref{fig:heat_map_mg1d,fig:heat_map_mg1d_4}, where $\log(\eta) / \beta$ approaches zero as the temperature decreases.
Thus, we conclude that the negative sign problem has been resolved for $J_3 > 2 \ \cup \ J_3 < J_2$, which includes the results in Ref.~\onlinecite{Nakamura1997_pv}.

\subsubsection{Bilinear-Biquadratic Chain}
Next, we consider the $S = 1$ bilinear-biquadratic (BLBQ) chain model.
The Hamiltonian encapsulates a range of significant models, with the parameter 
$\alpha$ tuning the system through various quantum states from the nearest-neighbor antiferromagnetic Heisenberg chain ($\alpha = 0$) to the AKLT model ($\alpha = 1/3$) and even to the intriguing SU$(3)$ chain, which can be solved exactly by the Bethe ansatz~\cite{Sutherland1975}, at $\alpha = 1$.
The Hamiltonian of the $S = 1$ BLBQ chain is defined as
\begin{equation}
	G = \sum_{i} g_{i,i+1},
\end{equation}
where the local interaction term $g_{i,i+1}$ is given by
\begin{equation}
  \label{eq:blbq_model}
	g_{i,i+1} = -\mathbf{S}_i \cdot \mathbf{S}_{i+1} - \alpha [{(\mathbf{S}_i \cdot \mathbf{S}_{i+1})}^2 - 1].
\end{equation}
Here, $\mathbf{S}$ represents the $S = 1$ spin operator.

\begin{figure}[tbp]
	\centering
  \includegraphics[clip,width=0.9\columnwidth]{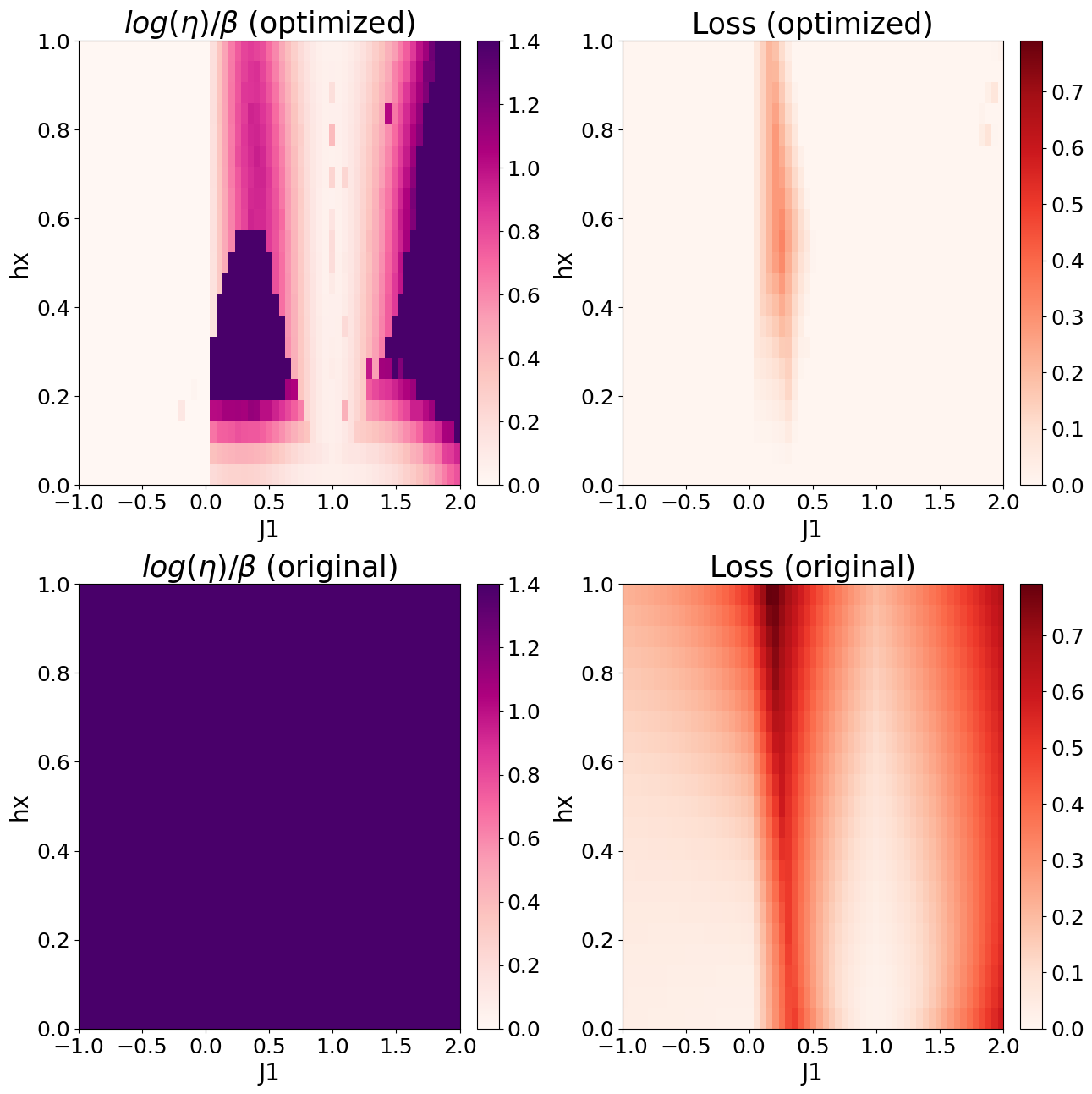}%
	\caption{
        Gap negativity $\eta^\prime \approx \log(\eta) / \beta$ (left) and L1 adaptive loss function (right) before (bottom) and after (top) the optimization for the BLBQ model at $\beta = 4$.
        Simulation is done for $L = 10$ with open boundary conditions.
	Color gradients indicate the magnitude of $\log(\eta) / \beta$ or the L1 adaptive loss function: 
	The lighter the color, the better.
    }\label{fig:heat_map_blbq}
\end{figure}

In our simulation, we employ cell type~(a) in \cref{fig:unit_cells}.
Building upon the original definition of the BLBQ model, we also examine the case where a finite transverse magnetic field $h_x$ is applied.
Typically, the transverse magnetic field is known to induce negative signs.
Since we know from the previous work that local unitary transformation can remove the negative sign, we can expect our method to also work well at the AKLT point and surrounding region ($\alpha < 1$).
The heat map of the negativity and loss function before and after the optimization is depicted in \cref{fig:heat_map_blbq}.
First, as analytically predicted, the negative sign is completely removed for $\alpha < 1$, as illustrated in the $h_x = 0$ section of \cref{fig:heat_map_blbq}.
While this result aligns perfectly with expectations, it does come as a bit of a surprise to us.
The reason is that the local basis transformation theoretically identified in Ref.~\onlinecite{Okunishi2013} is a unitary matrix, including complex elements, whereas, in the present study, we explore orthogonal matrices only.
Furthermore, we empirically confirm that our algorithm can get rid of the negative signs due to the transverse magnetic field, which can also be seen in regions $\alpha < 1$ with $h_x > 0$ in \cref{fig:heat_map_blbq}.
This finding demonstrates that our method can be applied to a broader range of models than just the frustration-free system.

\subsubsection{Shastry-Sutherland Model}

\begin{figure}[tbp]
	\centering
	\includegraphics[clip,width=0.5\columnwidth]{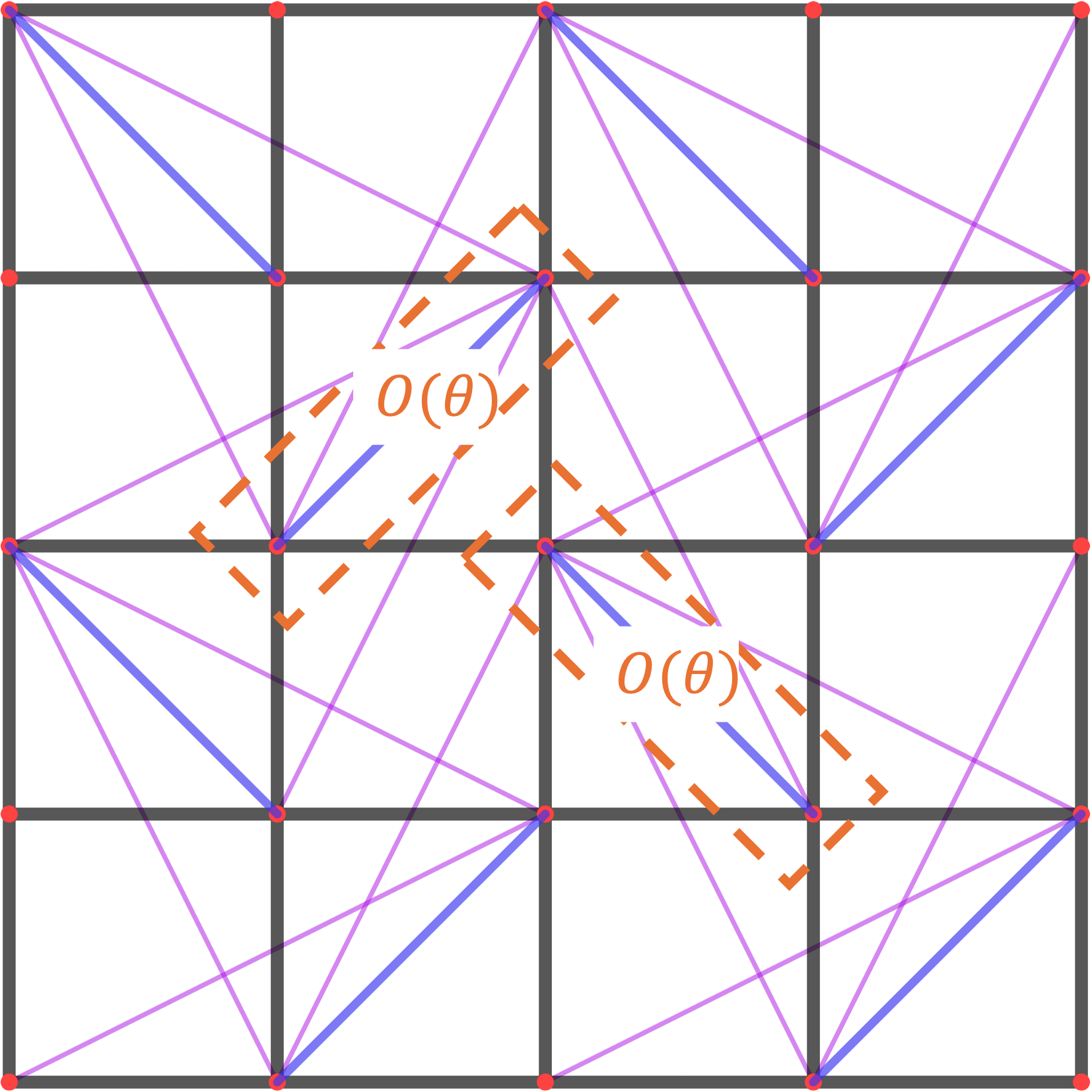}%
	\caption{
        Shastry-Sutherland model defined on the 2D square lattice.
	  The black, pink, and blue edges denote the Heisenberg interaction with coupling constant $J_d$, $J_1$, and $J_2$, respectively.
	  The cells for two-site basis transformation are denoted by dashed orange rectangles.
	}\label{fig:unit_cell_ss2d}
  \end{figure}

\begin{figure}[tbp]
  \centering
  \includegraphics[width=0.4\textwidth]{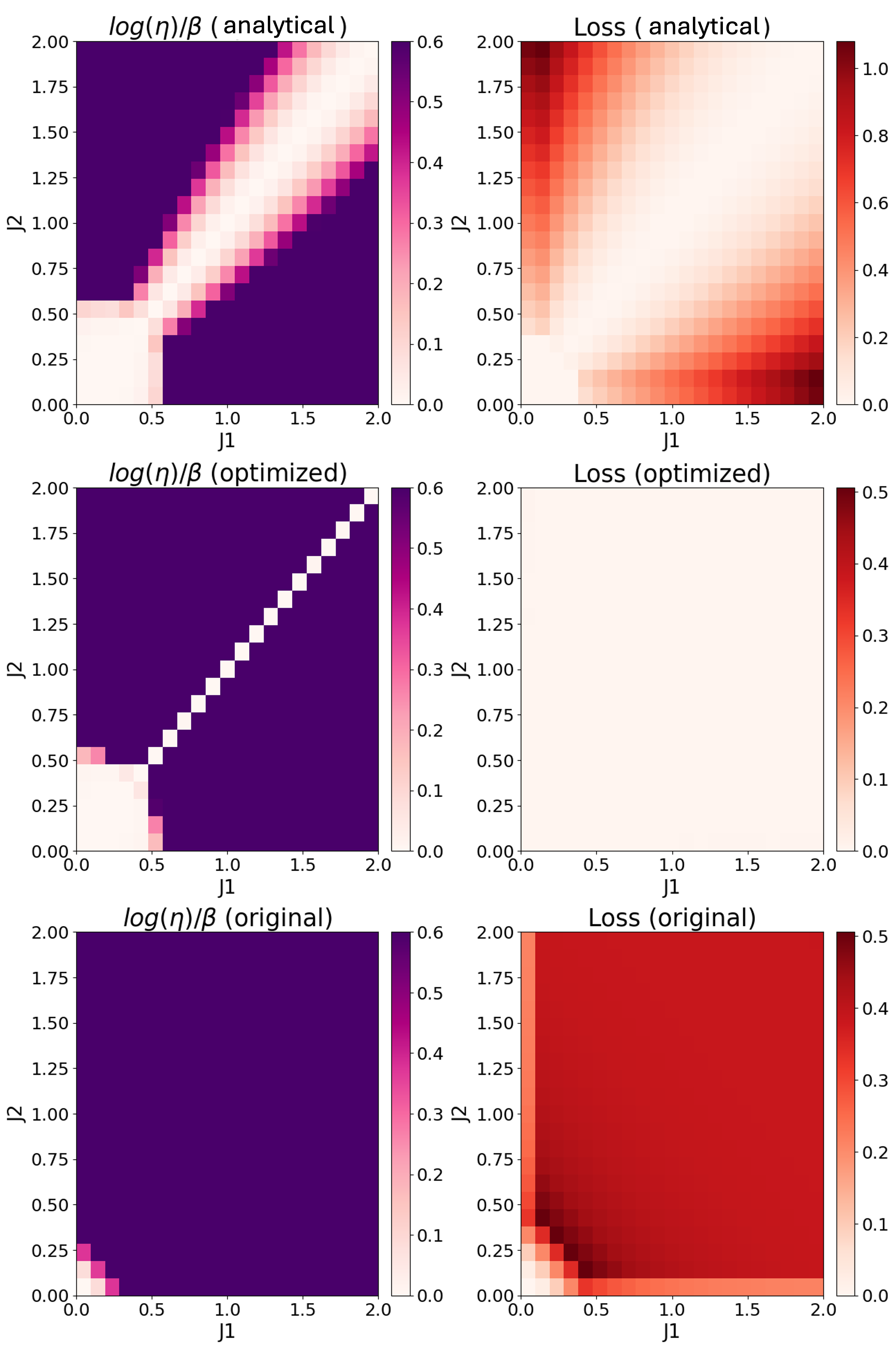}
  \caption{
        Gap negativity $\eta^\prime \approx \log(\eta) / \beta$ (left) and L1 adaptive loss function (right) before (bottom) and after (middle) the optimization for the Shastry-Sutherland model at $\beta = 4$ with $N = 4 \times 4$.
        For comparison, the results using the dimer-product basis are also presented in the top row.
  }\label{fig:heat_map_ss2d}
\end{figure}

As a final example of the frustration-free models, we examine the Shastry-Sutherland model,  a 2D frustrated spin model with known negative-sign-free local basis transformation~\cite{Miyahara1999,Wessel2018}. 
The lattice structure and the cell for the local basis transformation are shown in \cref{fig:unit_cell_ss2d}.
This model exhibits a wide variety of quantum phases, including dimer-singlet, spin liquid, Néel order, and plaquette-singlet~\cite{Koga2000_na,Yang2021_zg}, making it one of the most important examples in 2D frustrated systems.
It is known that, in the parameter region where $J_1 / J_d$ and $J_2 / J_d$ are small, the system becomes frustration-free, and the ground state is a unique dimer-singlet product state.
We choose the cells for the local basis transformation as shown in \cref{fig:unit_cell_ss2d}.
As seen in \cref{fig:heat_map_ss2d}, for the parameter region with smaller $J_1 / J_d$ and $J_2 / J_d$, we observe the negativity is completely removed.
This is an expected result as the system becomes frustration-free around this parameter region.
Also, we confirm that the parameter of $J_1 = J_2$ does not suffer from the negative sign problem with the analytical dimer basis as well as with our optimized basis.
The only apparent difference between the analytical solution (dimer basis) and our optimized basis is that the region deviated from the diagonal line $J_1 = J_2$.
In the dimer basis, the negativity remains small in a wide region around $J_1 = J_2$.
In our optimized basis, however, the negativity becomes unity only when $J_1 = J_2$.

\begin{figure}[tp]
  \centering
  \includegraphics[width=0.5\textwidth]{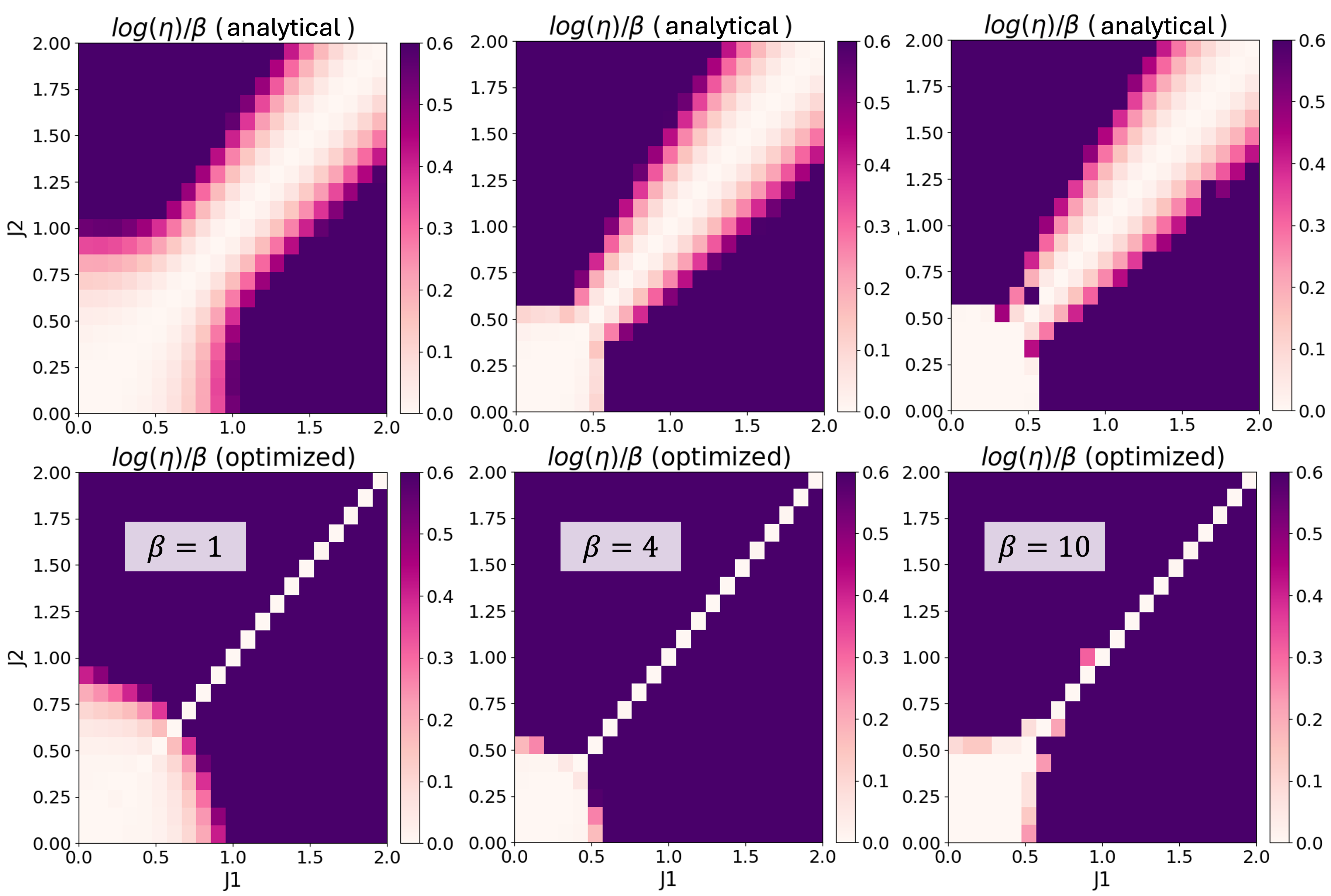}
  \caption{
        Gap negativity $\eta^\prime \approx \log(\eta) / \beta$, of the Shastry-Sutherland model at different temperatures.
        The first row is the result using the dimer-product basis, and the second row is the result using the optimized basis.
		Note that even for the analytically derived basis, the parameter region, except for the diagonal line and the bottom-left square exhibits the severe negative sign problem at low temperatures.
		For example, $\log(\eta) / \beta = 0.345$ at $J_1 = 0.8, J_2 = 1.0$, and this is about $\eta^{-1} = \expval{S} = 0.0317$ at $\beta = 10$, which requires around 1,000 times more Monte Carlo steps to achieve the same accuracy as the one without negative sign problem.
  }\label{fig:heat_map_ss2d_0.1}
\end{figure}

The observed difference between the dimer-product basis and the optimized basis can be attributed to the nature of our loss function, which is not guaranteed to approximate the average sign when the system is not frustration-free. 
Conversely, this result can also be interpreted as evidence that even a small deviation from the diagonal line 
$J_1 = J_2$ causes the Shustry-Sutherland model to deviate significantly from the frustration-free property. 
Nevertheless, it is important to note that the analytically derived basis also suffers from the sign problem at low temperatures in the parameter region slightly off the diagonal line, as shown in \cref{fig:heat_map_ss2d_0.1}. 
This means that both the analytically derived basis and our optimized basis behave similarly at low temperatures.

\begin{figure}[tbp]
	\centering
	\subfloat[Simulation at $\beta=0.25, h_x = 0$ ]{%
	  \includegraphics[clip,width=0.9\columnwidth]{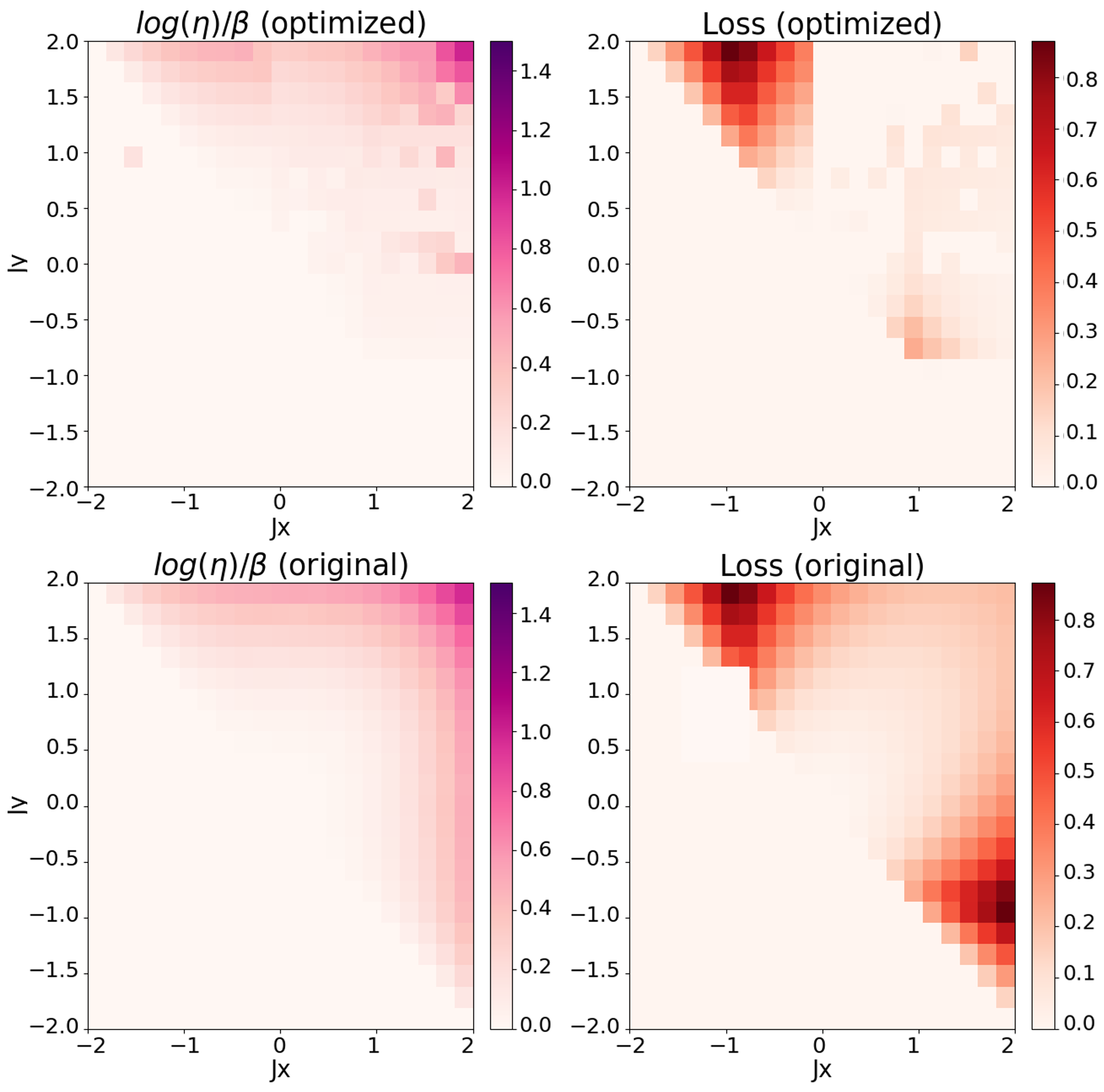}%
	}\\
	\subfloat[Simulation of $\beta=0.25, h_x = 0.5$]{%
	  \includegraphics[clip,width=0.9\columnwidth]{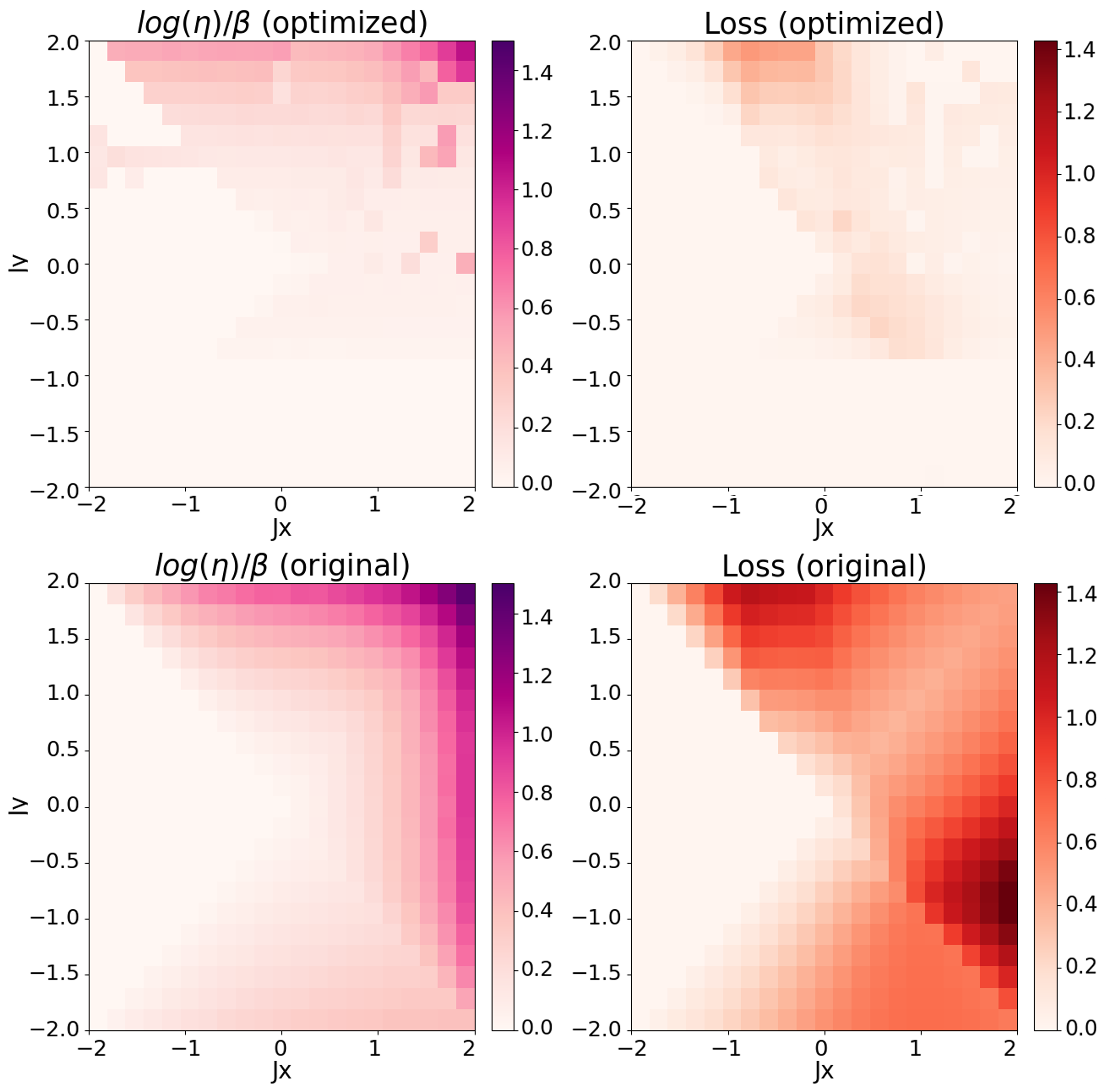}%
	}\\
	\centering
	\subfloat[Simulation at $\beta=0.5, h_x = 0$ ]{%
	  \includegraphics[clip,width=0.9\columnwidth]{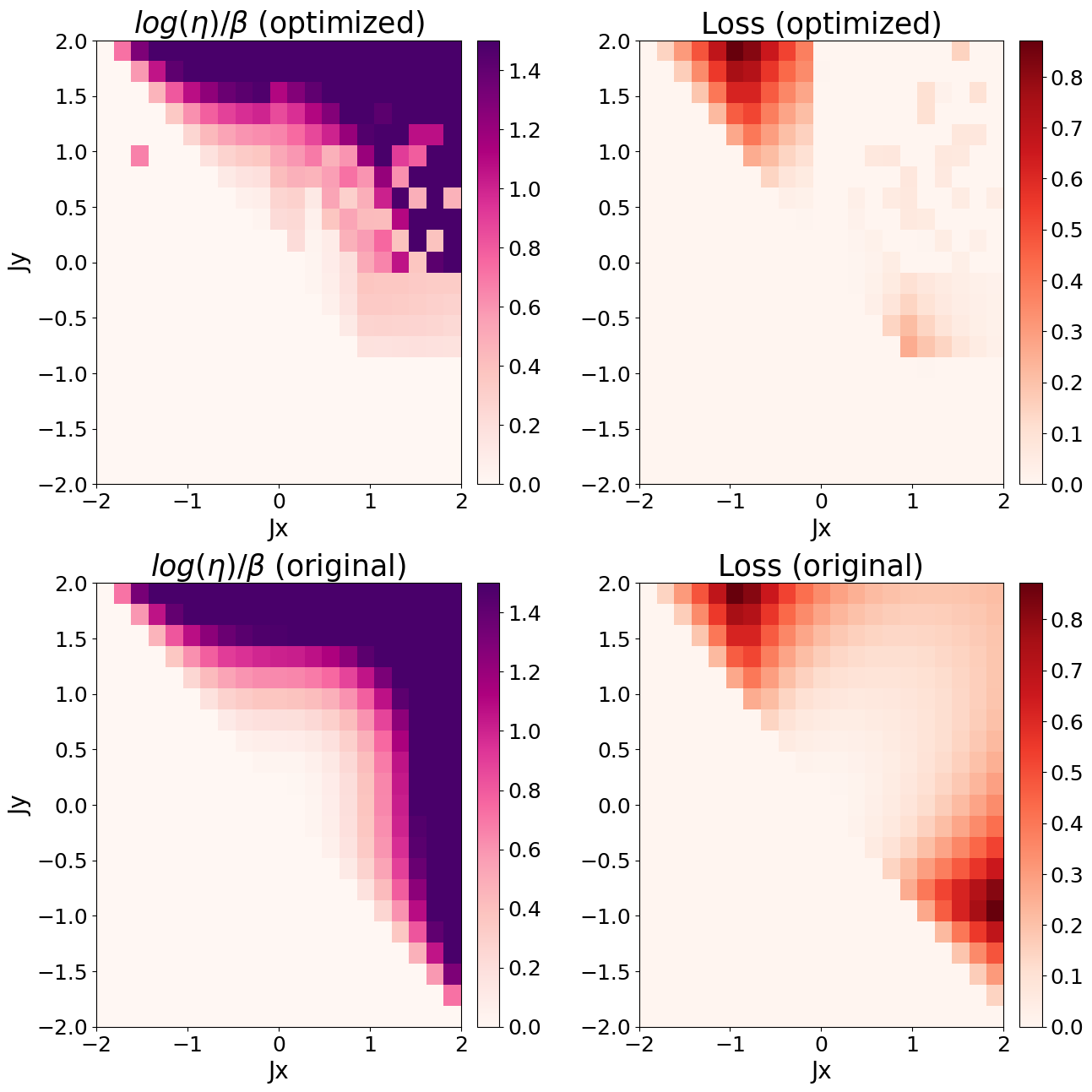}%
	}\\
  \end{figure}
  \begin{figure}[tbp]
	\ContinuedFloat
  
	\subfloat[Simulation at $\beta=0.5, h_x = 0.5$]{%
	  \includegraphics[clip,width=0.9\columnwidth]{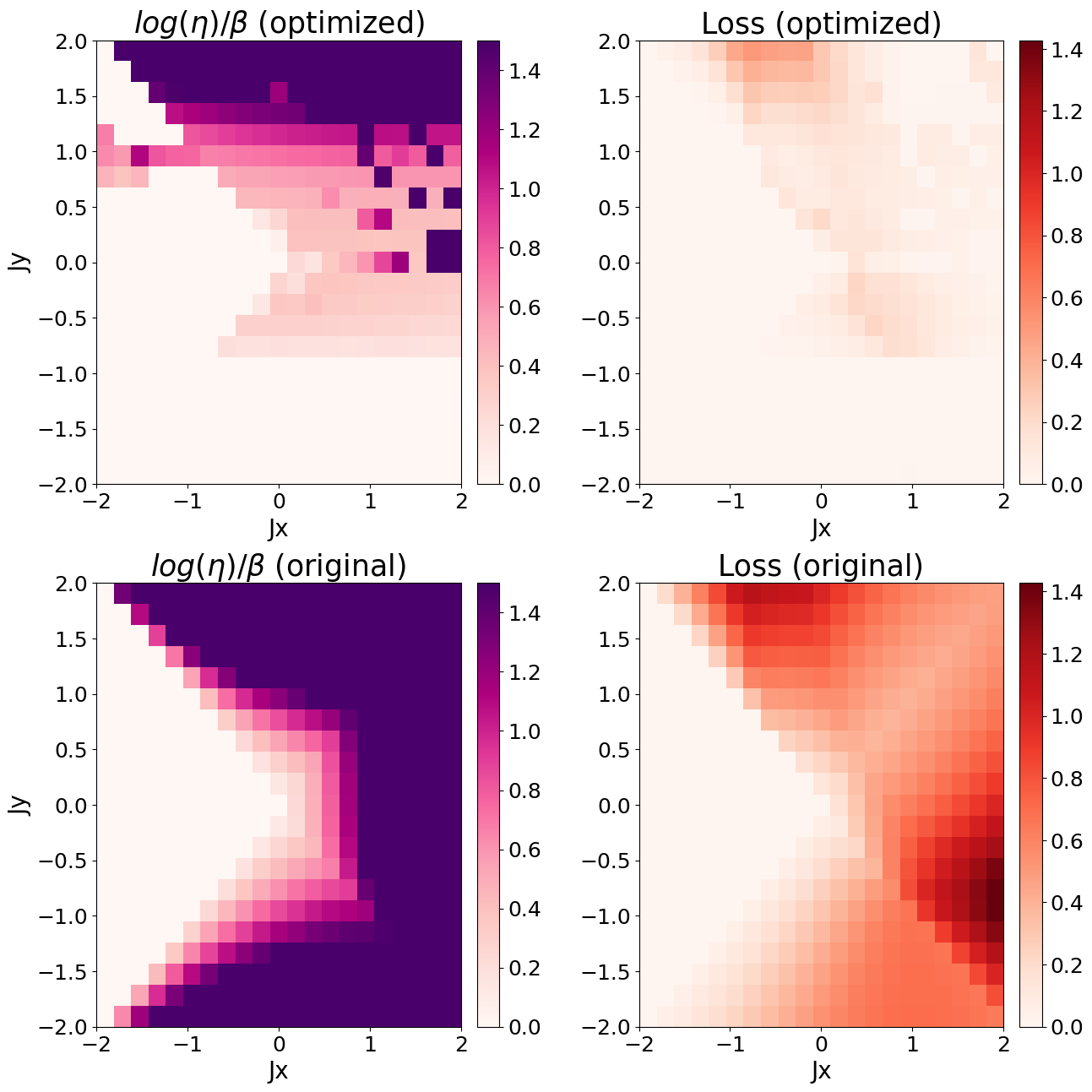}%
	}\\
	\caption{
      Gap negativity $\eta^\prime \approx \log(\eta) / \beta$ (left) and L1 adaptive loss function (right) before (bottom) and after (top) the optimization for the anisotoropic kagome Heisenberg model at $(\beta, h_x) = (0.25, 0)$ (a), $(0.25, 0.5)$ (b), $(0.5, 0)$ (c), and $(0.5, 0.5)$ (d).
	  The coupling constant $J_z$ is fixed to 1, with the vertical axis representing $J_y$ and the horizontal axis representing $J_x$, both ranging from -2 to 2.
	  Each heatmap should be interpreted in the same way as in the other figures.
	  The system size is $N = 5 \times 5$.
	  We confirmed that $N=4 \times 4$ and $6 \times 6$ system sizes give similar results.
	}\label{fig:heat_map_kagome}
  \end{figure}

  \begin{figure}[tp]
	  \centering
	  \includegraphics[clip,width=0.7\columnwidth]{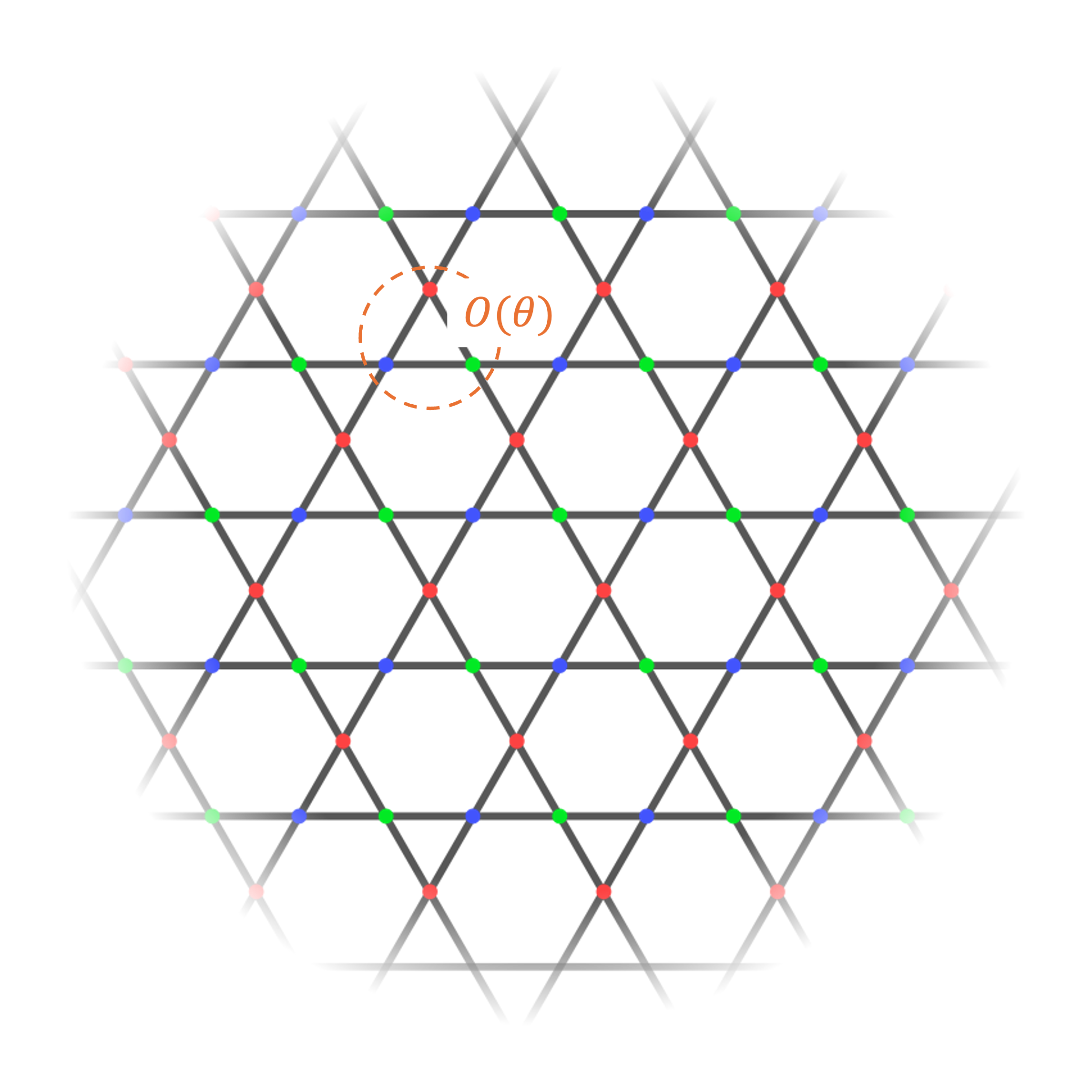}%
	  \caption{
		  Lattice structure of the kagome Heisenberg model and cell for three-site basis transformation.
		  In the figure, we combine the three sites into a single cell.
	  }\label{fig:kagome}
  \end{figure}
  
\subsection{Kagome Heisenberg Model}
\label{sec:kagome}

Lastly, we apply our method to the Heisenberg model on the kagome lattice as a promising candidate 
to exhibit exotic quantum-dynamical behavior such as spin liquid~\cite{Hermele2008_vs}.
The precise nature of its ground state, as well as finite temperature properties, however, remains actively debated, with proposed possibilities including valence bond crystals ground states ~\cite{Nikolic2003_zy}.
From a computational perspective, the kagome Heisenberg model poses significant difficulties due to frustrating interactions and its intrinsic 2D structure.
While some techniques like exact diagonalization are limited to small system sizes, and density matrix renormalization group studies have relied on the quasi-1D approximation of cylinders or strips, QMC stands out in its potential ability 
to simulate large 2D kagome lattices.
Unfortunately, the presence of frustrating antiferromagnetic interactions makes the QMC simulation of the kagome Heisenberg model a challenging task~\cite{Nakano2010-bz}.

There is previous research on alleviating the negative sign problem for the kagome Heisenberg model.
In Ref.~\onlinecite{DEmidio2020_yx}, we confirmed that with a strong transverse magnetic field ($h_x/J \sim 4$), the negative sign disappears at low temperatures, without any basis transformation.
This can be understood from the fact that for any model, the negative sign vanishes in limits where the transverse field becomes dominant. While an interesting result, especially showing the phenomenon of the sign disappearing only at low temperatures, as seen in the Shastry-Sutherland model, this approach has greatly limited applicability.
In the present numerical experiments, there is no transverse magnetic field, or only a weak field is applied.
The cell is taken to contain three sites as shown in \cref{fig:kagome}.
We analyze the cases where the interaction is anisotropic: $J_z$ is fixed to unity, while $J_x$ and $J_y$ are changed from -2 to 2.
We confirm that the negative sign is alleviated in certain parameter region \cref{fig:heat_map_kagome}.
These results also lead to an intriguing conjecture: our method almost entirely eliminates the negative sign caused by the transverse magnetic field.
In fact, by comparing (a) and (c) with (b) and (d) in \cref{fig:heat_map_kagome}, we can see that the effect of the magnetic field on the negative sign is removed to a large extent.
Another noteworthy observation is that, without a transverse magnetic field, the model is invariant under the exchange between $J_x$ and $J_y$.
However, in our results, this symmetry is broken, as seen clearly in \cref{fig:heat_map_kagome}.
This issue, which is explored in more detail in the next section, arises because we permit only orthogonal transformations as the local basis transformations.

\begin{figure}[tbp]
	\centering
	\includegraphics[clip,width=1.0\columnwidth]{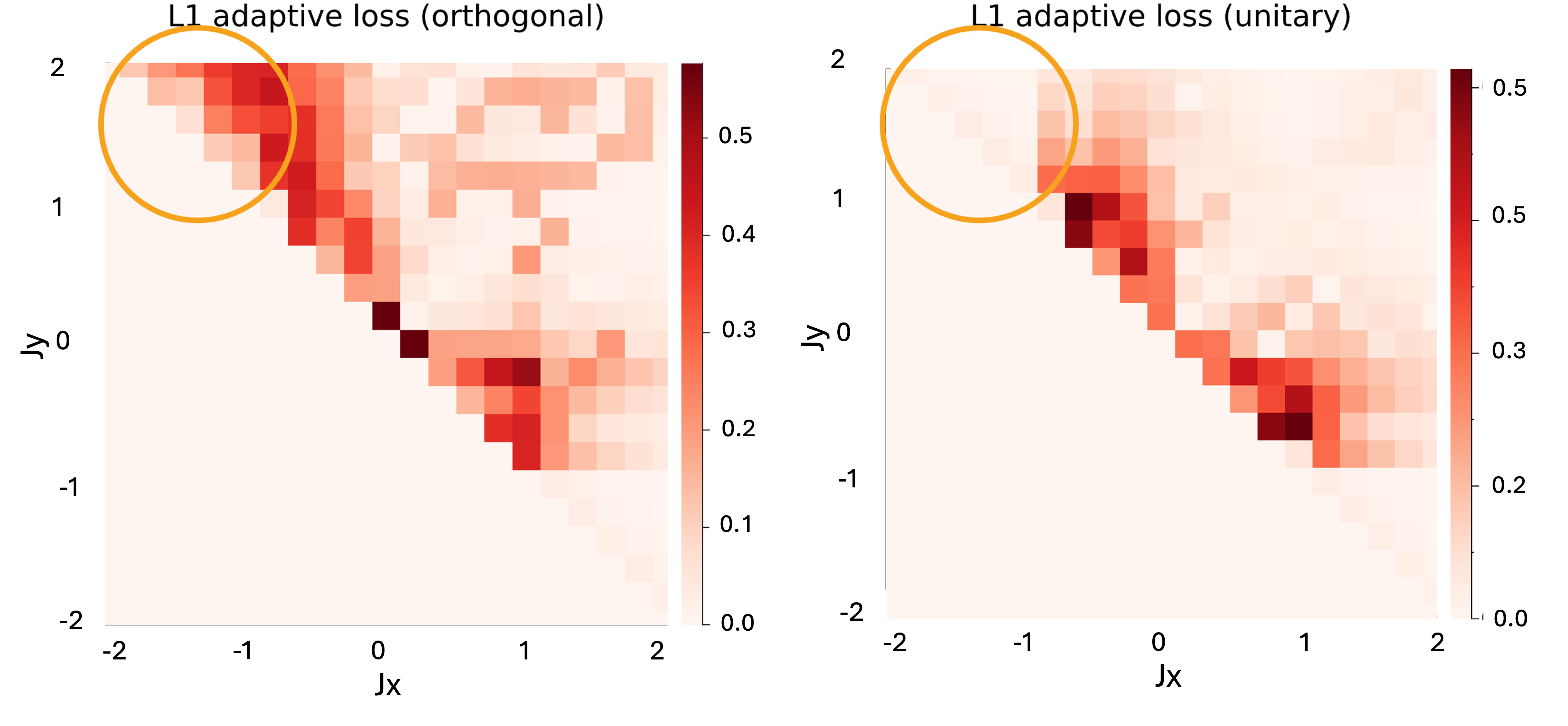}%
	\caption{
		Comparison of L1 adaptive loss function with unitary and orthogonal transformations 
		after optimization for the kagome Heisenberg Model at $J_z = 1$ and $h_x = 0$.
		The left figure is the result of the orthogonal transformation, and the right figure is the result of the unitary transformation.
		Orange circle indicate the region where the negativity is removed only by the unitary transformation.
	}\label{fig:unitary_vs_orthogonal}
\end{figure}

\section{Orthogonal vs Unitary Transformations}\label{sec:orth_vs_unitary}

The numerical experiments in the previous sections as well as in previous studies~\cite{Hangleiter2020, Levy2021, Marvian2019_hh}, only orthogonal transformations were considered as local transformations.
However, it is worth noting that in QMC, the weights can be complex values, which allows for the use of unitary transformations.
Here, we explore the possibility that utilizing unitary transformations could further mitigate the negative sign problem.
We demonstrate it by comparing the results of unitary and orthogonal transformations applied to the kagome Heisenberg Model.
As seen in \cref{fig:heat_map_kagome} in \cref{sec:kagome}, if we optimize within orthogonal transformations, the inherent symmetry under the exchange of $J_x$ and $J_y$ was broken.
However, when allowing for transformations that include unitary transformations, the symmetry is restored as shown in \cref{fig:unitary_vs_orthogonal}.
This is a highly intriguing result, indicating that even though the original Hamiltonian is a real symmetric matrix,
using unitary matrices instead of orthogonal ones can break the real symmetry to 
potentially mitigate the negative sign problem even more.

\section{Conclusion}\label{sec:conclusion}
In the present study,
we first introduced a unified metric specifically designed to represent the severity of the negative sign problem.
This metric not only facilitates a cohesive treatment of both reweighting and basis transformation methods,
but it also serves as a definitive indicator of the severity of the negative sign.
All other metrics employed to address the negative sign problem in the future 
should ideally be derived from this foundational measure,
ensuring a consistent and comprehensive approach to quantifying and mitigating the issue.
A key finding from our investigation was the realization that employing the stoquastic form of Hamiltonian $G^+$ as a virtual Hamiltonian $G_\text{v}$ 
in reweighting is always the optimal selection in terms of negativity.

Particularly,
we formulated the basis transformation methods as an optimization problem,
and focused mainly on the L1 adaptive loss function we devised under the assumption of the frustration-free property.
This property is a commonly observed characteristic 
in quantum spin models where local basis transformation methods 
are analytically known to function effectively.
We then applied this optimized approach to several quantum spin models.
Our method demonstrated excellent performance,
single-handedly reproducing almost all previously discovered 
basis transformations found in various studies.
The implications of this research are substantial,
offering a new lens through which the negative sign problem can be approached and potentially mitigated,
even in quantum spin systems where analytical solutions are unknown.

However, the study also revealed certain limitations.
Although our method performed more than expected in some models
with no relation to the frustration-free property,
the method does not guarantee the improvement of the negative sign problem and shows potential inaccuracies 
when applied outside the frustration-free domain.
Additionally, we found that the method exhibited 
lower improvement rates in 2D systems compared to its great success in 1D systems.
These limitations likely stem from the approximate nature of the L1 adaptive loss function,
which may not fully encapsulate the complexities of lattice structures.

\begin{acknowledgments}
The authors would like to thank Tsuyoshi Okubo and Hidemaro Suwa
for discussions and comments. 
This work was supported by the Center of Innovation for Sustainable Quantum AI,
JST Grant Number JPMJPF2221, JSPS KAKENHI Grant Numbers JP20H01824 and JP24K00543,
and JSPS KAKENHI Grants No. JP24KJ0892.

\end{acknowledgments}

\bibliographystyle{unsrt}
\bibliography{main} 
\end{document}